\documentclass[a4paper,11pt]{article}
\pdfoutput=1 
\usepackage{jinstpub}
\usepackage{graphicx} 
\usepackage{url}
\usepackage{multirow}
\usepackage{amsmath}

\usepackage[dvipsnames]{xcolor}

\title{\boldmath A density-based clustering algorithm for the \\CYGNO data analysis}

\author[a,b]{E. Baracchini,} 
\author[c]{L. Benussi,}
\author[c]{S. Bianco,}
\author[c]{C. Capoccia,} 
\author[c,d]{M. Caponero,}
\author[e,f]{G. Cavoto,}
\author[a,b]{A. Cortez,}
\author[g]{I. A. Costa,}
\author[e]{E. Di Marco,}
\author[e]{G. D'Imperio,}
\author[a,b]{G. Dho,}
\author[e]{F. Iacoangeli,}
\author[c]{G. Maccarrone,}
\author[e,h]{M. Marafini,}
\author[c]{G. Mazzitelli,}
\author[e,f]{A. Messina,}
\author[g]{R. A. Nobrega,}
\author[c]{A. Orlandi,}
\author[c]{E. Paoletti,}
\author[c]{L. Passamonti,}
\author[i,j]{F. Petrucci,}
\author[c]{D. Piccolo,}
\author[c]{D. Pierluigi,}
\author[e]{D. Pinci}
\author[e]{F. Renga,}
\author[c]{F. Rosatelli,}
\author[c]{A. Russo,}
\author[c,k]{G. Saviano,}
\author[c]{and S. Tomassini}

\affiliation[a]{Gran~Sasso~Science~Institute,\\ L'Aquila, I-67100, Italy}
\affiliation[b]{Istituto Nazionale di Fisica Nucleare,\\ Laboratori Nazionali del Gran Sasso, Assergi, Italy}
\affiliation[c]{Istituto Nazionale di Fisica Nucleare ,\\  Laboratori Nazionali di Frascati, I-00044, Italy}
\affiliation[d]{ENEA Centro Ricerche Frascati, Frascati, Italy}
\affiliation[e]{Istituto~Nazionale~di~Fisica~Nucleare,\\ Sezione di Roma, I-00185, Italy}
\affiliation[f]{Dipartimento di Fisica Sapienza Universit\`a di Roma, I-00185, Italy} 
\affiliation[g]{Universidade Federal de Juiz de Fora, Juiz de Fora, Brasil}
\affiliation[h]{Museo Storico della Fisica e Centro Studi e Ricerche "Enrico Fermi",\\ Piazza del Viminale 1, Roma, I-00184, Italy}
\affiliation[i]{Dipartimento di Matematica e Fisica, Universit\`a Roma TRE, Roma, Italy}
\affiliation[j]{Istituto Nazionale di Fisica Nucleare, Sezione di Roma TRE, Roma, Italy}
\affiliation[k]{Dipartimento di Ingegneria Chimica, Materiali e Ambiente, Sapienza Universit\`a di Roma, Roma, Italy}

\emailAdd{igor.abritta@engenharia.ufjf.br}

\date{March 2020}

\abstract{
Time Projection Chambers (TPCs) working in combination with Gas Electron Multipliers (GEMs) produce a very  sensitive detector capable of observing low energy events. This is achieved by capturing photons generated during the GEM electron multiplication process by means of a high-resolution camera.
The CYGNO experiment has recently developed a TPC Triple GEM detector coupled to a low noise and high spatial resolution CMOS sensor.
For the image analysis, an algorithm based on an adapted version of the well-known DBSCAN was implemented, called iDBSCAN.
In this paper a description of the iDBSCAN algorithm is given, including test and validation of its parameters, and a comparison with DBSCAN itself and a widely used algorithm known as Nearest Neighbor Clustering (NNC).
The results show that the adapted version of DBSCAN is capable of providing full signal detection efficiency and very good energy resolution while improving the detector background rejection.
}

\begin{document}
\maketitle
\flushbottom

\section*{Introduction}

Clustering analysis is a widely used unsupervised technique to organize datasets into groups based on their similarities.
One of the most known algorithms is the so-called Density-Based Spatial Clustering of Applications with Noise (DBSCAN) \cite{dbscan1996}.
Given a set of elements distributed over a hyper-plane, DBSCAN seeks for areas of high density to form clusters. Such density is calculated considering the number of elements within a pre-defined hyper-sphere.
The generalization power of DBSCAN and its simplicity, which make it a very attractive algorithm, can be understood in terms of its two parameters: the radius of the hyper-sphere ($\epsilon$), which is applied over each element to count the number of neighboring elements around it, and the minimum number of points inside each hyper-sphere ($N_{min}$), used to decide if those elements should make up a cluster.
To fulfill the needs of the CYGNO experiment, a detector-specific algorithm, based on DBSCAN, has been developed.
Within the context of the experiment, a detection apparatus composed of an optical readout system based on a high-resolution and low noise CMOS sensor capable of providing track images produced by interacting particles with release energies in the range of a few keV has been developed \cite{bib:ref1,bib:ref2,bib:ref3,bib:ref4,bib:nim_orange1,bib:jinst_orange2}.
This modified version of DBSCAN, called intensity-based DBSCAN or simply iDBSCAN, has shown to be able to improve detector performance when compared to the previously used algorithm based on the Nearest Neighbor Clustering (NNC) technique \cite{bib:fe55}. This paper proposes a study on the impact of iDBSCAN when compared to NNC and DBSCAN on two crucial detector's parameters, background rejection and energy resolution, measured in the energy range of a few keV. For such, low energy particles (5.9 keV photons) produced by a $^{55}$Fe radioactive source, background from natural radioactivity and data with electronics noise only were employed. 

\section{Experimental setup}
\label{sec:expSetup}

\subsection{LEMOn detector}

LEMOn (Large Elliptical MOdule)~\cite{bib:ieee17} is the most recent CYGNO experiment's prototype. Its core consists of a 7 liter active drift volume surrounded by an elliptical field cage ($\rm 20 \times 20 \times 24~cm^3$) and a $\rm 20 \times 24~cm^2$ Triple GEM structure whose produced photons are readout by an Orca Flash 4 CMOS-based camera~\cite{ORCAcamera} placed at a distance of $\rm 52.5~cm$ (i.e. 21 Focal Length, FL).
More details are given in Ref. \cite{bib:fe55, bib:eps}.
The drift chamber was filled with a $\rm He/CF_4$ gas mixture in the proportion of $60/40$ and a $^{55}$Fe source with an activity of about 740 MBq was used.
For operation, electric fields are applied to the TPC drift volume and between the GEMs. They are called drift field ($E_d$) and transfer field ($E_t$) respectively.
The typical operating conditions of the detector, as used in this work, are: $E_d$ = 500 V/cm, $E_t$ = 2.5 kV/cm, and a voltage difference across the GEM sides ($ V\rm_{GEM}$) of 460 V.

\begin{figure}[ht]
\centering
\includegraphics[width=0.85\textwidth]{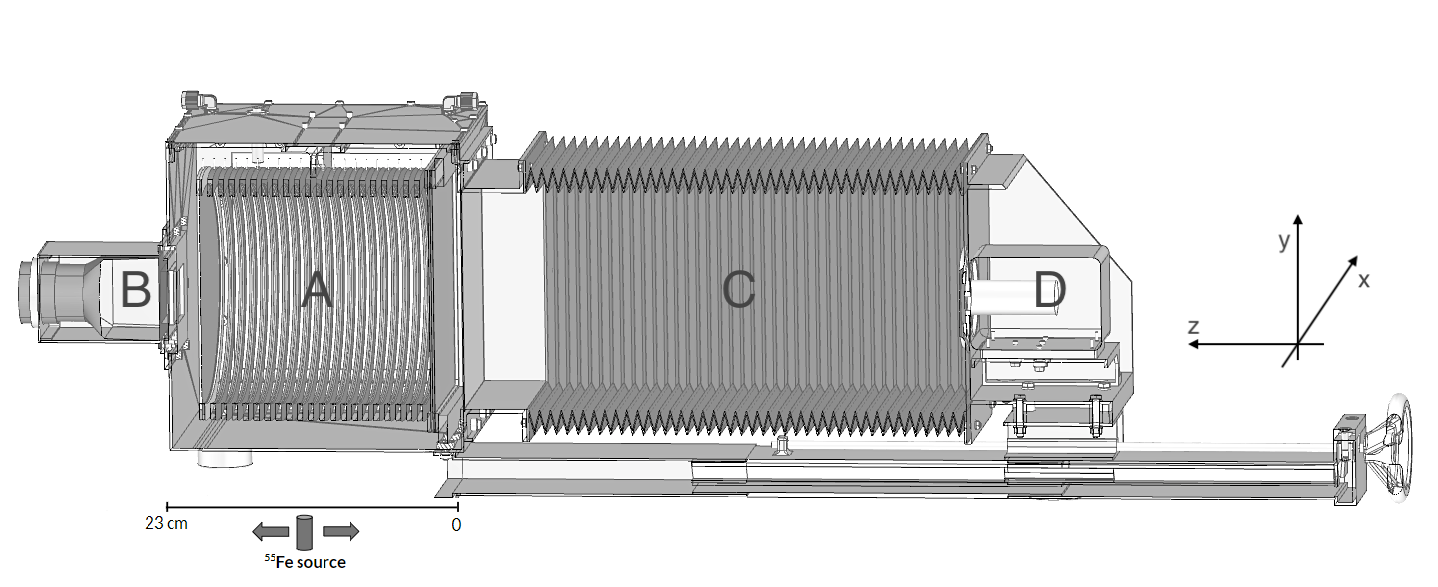}
\caption{Drawing of the experimental setup. In particular, the elliptical field cage close on one side by the triple-GEM structure and on the other side by the semitransparent cathode (A), the PMT (B), the adaptable bellow (C) and the CMOS camera with its lens (D) are visible. The sliding external $^{55}$Fe source, positioned close to the TPC is also drawn.} \label{fig_lemon_1}
\end{figure}

\subsection{Acquisition runs}
\label{sec:acrRuns}

Data were acquired using auto-trigger mode. For the proposed study presented in this document, three different acquisition datasets were used, as listed below:

\begin{itemize}
    \item Electronic noise (EN) dataset: produced by lowering down $V\rm _{GEM}$ to 260 V, a value where the multiplication process is forbidden (6478 images recorded);
    \item Natural radioactivity (NRAD) dataset (composed of cosmic rays and environmental radioactivity): produced by exposing the camera lens and turning on the detector power supplies and raising $V\rm _{GEM}$ to the nominal value of 460 V to allow charge multiplication and secondary light emission during this process (864 images recorded);
    \item Electron Recoils (ER) dataset: the same as the previous item but placing a $^{55}$Fe source near to the detector drift volume (864 images recorded).
\end{itemize}

\subsection{Detector expected signals}

\begin{figure}[!ht]
\centering
\includegraphics[width=0.7\textwidth]{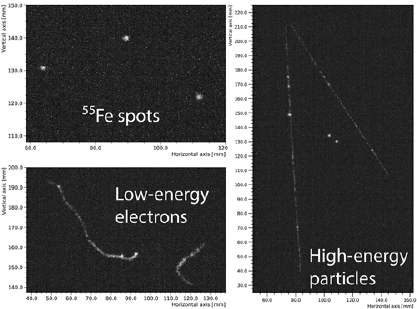}
\caption{Examples of signals that can occur using the described configuration.} \label{fig_example}
\end{figure}

Based on the acquisition datasets defined in section \ref{sec:acrRuns}, particles interacting with the detector gas can have two distinct origins: $^{55}$Fe source and natural radioactivity. The former releases 5.9 keV photons which produce round spots on the image while the latter can be composed of few different particles as photons, electrons and muons. 
Typical signals are shown in  Fig.~\ref{fig_example}:
three interactions of $^{55}$Fe photons in the left top image; two low-energy electrons in the left bottom image; and two high-energy particles (likely to be cosmic ray muons) and, between them, two interactions of $^{55}$Fe photons in the right image.

In this work, the signals of interest are those generated by the 5.9 keV photons, which are used to assess the impact of the proposed clustering algorithms on the detector characteristics, focusing mainly on its energy resolution and background-events rejection performance in the energy range of few keV.

\section{Data analysis flow}
\label{sec:daq}

\subsection{Data structure}

\begin{figure}[ht]
\centering
\includegraphics[width=0.45\textwidth]{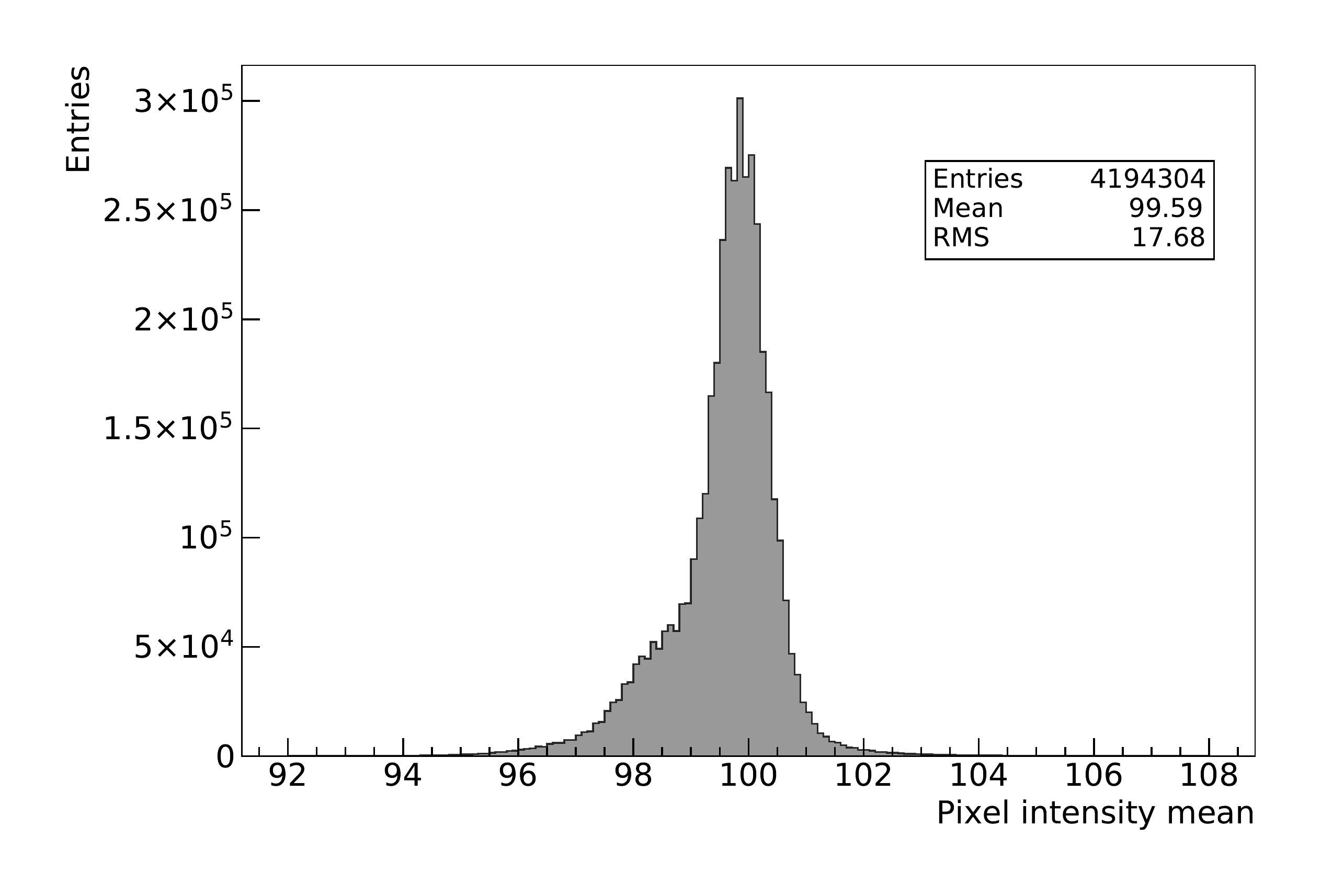}
\includegraphics[width=0.45\textwidth]{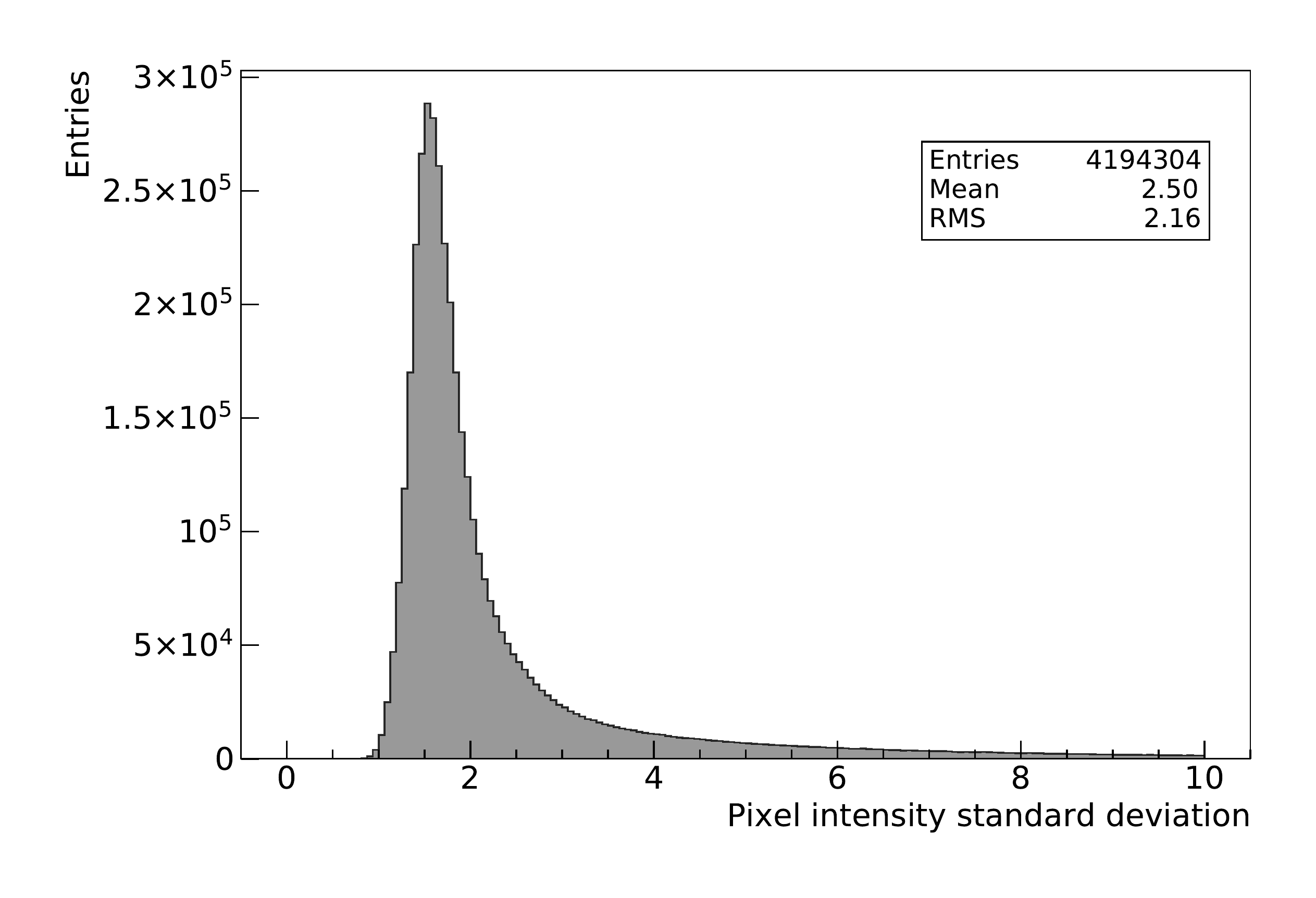}
\caption{Mean and standard deviation distributions of the sensor's pixels noise.}
\label{fig:sensor_noise}
\end{figure}

The acquisition system provides images with 2048 $\times$ 2048 pixels captured by the Orca Flash 4 V3 CMOS sensor. The photo sensor has a sensitive area of 13312 $\times$ 13312 $\mu$m$^2$ and each pixel has a size of 6.5 $\mu$m $\times$ 6.5 $\mu$m.
The camera's exposure time was set to 40 ms and it covers an area of 26 $\times$ 26 cm$^2$ in relation to the plane of the last layer of the GEM detector. Each pixel provides a response, here called intensity, proportional to the number of collected photons \cite{bib:nim_orange1} added to a baseline, also known as pedestal, which can be defined as the intensity value corresponding to zero photons. Specifically, the pedestal average value of the sensor is about 99 counts, however it can vary from pixel to pixel.
Additionally, the noise level is another important parameter that can vary from pixel to pixel.
Those effects can be seen in Fig. \ref{fig:sensor_noise}, which shows the mean and standard deviation distributions of the noise as computed for each pixel, produced with the EN dataset.
To account for such variations, both the pixel baseline ($\mu_i$) and its average noise ($\sigma_i$), calculated as the standard deviation of the pedestal distribution,  are estimated for every single pixel $i$ before running the event reconstruction procedure.

\subsection{Overview of the event reconstruction procedure}

\begin{figure}[ht]
\centering
\includegraphics[width=0.9\textwidth]{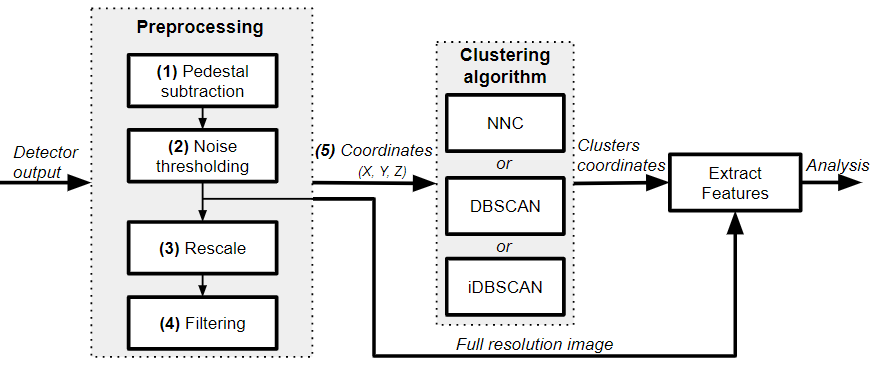}
\caption{Flowchart of the CYGNO's event-reconstruction algorithm.}
\label{fig_algo}
\end{figure}

The current CYGNO's event-reconstruction algorithm is represented in the flowchart shown in Fig.~\ref{fig_algo} and it is described below.

\begin{enumerate}
\addtocounter{enumi}{0}
    
    \item Pedestal subtraction is carried out pixel by pixel by subtracting $\mu_i$ from their original intensity values, generating new intensity values defined as $I_i$. 

    \item Lower and upper thresholds are applied to $I_i$.
    While the upper limit is set to $100$ counts, the lower limit is set to 1.3 times $\sigma_i$.
    The upper limit allows to remove pixels with a too large intensity, produced mainly due to leakage currents that go into sensor wells - also known as hot-pixels, while the lower limit was optimized and set to be just above noise level to ensure a good detection efficiency, but not too low in order not to overload the event-reconstruction algorithm with pixels dominated by noise. Pixels outside those limits have their intensities reset to zero.

    \item Images are then rescaled to 512 $\times$ 512 pixels, for CPU reasons, so that each $4 \times 4$ matrix, called macro-pixel, is assigned an intensity value corresponding to the average of the intensities $I_i$ of the 16 pixels occupying the same area of the sensor. 
    
    \item The rescaled image goes then through a filtering stage based on a $4 \times 4$ median filter that replaces a given macro-pixel intensity by the median of all macro-pixels in its neighborhood $w$, $g(x,y)$, as given by Equation \ref{eq10} \cite{filterStudy}, where $f(x,y)$ is the intensity of the macro-pixel $(x,y)$.
        \begin{equation}
             g(x,y) = median\{f(x,y), (x,y) \in w\}
             \label{eq10}
         \end{equation}
        
    Such filter is widely used in many applications due to its effective noise suppression capability and high computational efficiency~\cite{gonzalez2002digital}.
    Tests performed on the EN dataset (see section \ref{sec:acrRuns}) showed that this filter is able to reduce the number of noise pixels sent to the clustering algorithm by a factor of 3.07 $\pm$ 0.02.

    \item Finally, the coordinates (X, Y) and respective intensities (Z) of the pixels with non-zero $I_i$ values are sent to the clustering algorithm whose output is used to extract clusters' features such as integrated light, length and width, computed over the full-resolution image. Those features are then used to select events of interest. 
\end{enumerate}

In this work three features, extracted from the clusters, are used: 
\begin{itemize}
    
    \item Length and width:
    the full length of the major and minor axes along the two eigenvectors of the (X,Y) pixel matrix in the context of Principal Component Analysis \cite{jolliffe2002springer} are assigned as the length and width of the cluster, respectively.
    
    \item Cluster light: calculated as the sum of all the pixel $I_i$ intensities belonging to the cluster.
\end{itemize}

As mentioned before, prior to iDBSCAN, the CYGNO clustering algorithm was based on the widely employed NNC method. Basically it groups neighboring pixels that went through a selection similar to the one in step 3. A detector performance study using such method was presented in \cite{bib:fe55}.
To understand the advantages of using iDBSCAN, in addition to the comparison with NNC, the performance achieved with the DBSCAN algorithm will also be presented.

\subsection{The CYGNO intensity-based clustering algorithm}
\label{sec:dbscan}

\subsubsection{iDBSCAN}

As in many areas, in particle physics it is possible to insert a priori knowledge about the detection system and its data to improve the performance of the clustering task \cite{wagstaff2000clustering}. In this sense, a modification of DBSCAN \cite{scikit-learn} clustering algorithm was implemented, to better match the experimental conditions and data of the LEMOn detector.
As mentioned before, DBSCAN has only two parameters: $\epsilon$ and $N_{min}$. Whenever the number of neighboring elements inside a hyper-sphere reaches the $N_{min}$ value, the center element and all its neighbors are activated to start the formation of a cluster.
Then, the same process happens to all the neighboring elements in order to expand the starting cluster, to form a final cluster. This process is repeated to all the data elements.
To be applied to CYGNO, instead of using the number of elements as a parameter to decide if the elements inside a hyper-sphere make part of a cluster, the sum of their intensity values is used. Consequently, the $N_{min}$ becomes a parameter related to the total intensity within a hyper-sphere instead of to the number of elements.
Therefore, rather than having each pixel counted as a unit when computing the number of pixels inside a given hyper-sphere, each pixel counts $I_i$ times.
If the total intensity is equal or greater than a certain value ($N_{min}$), they are considered as making part of a cluster.
During the development of the iDBSCAN algorithm, many $\epsilon$ and $N_{min}$ values have been tested, leading us to converge to values around 5.8 and 30, respectively, which will be validated in section~\ref{idbscan_vali}.
Additionally, to make iDBSCAN more robust against electronic noise and intensity spikes, a cluster is required to have more than two macro-pixels, otherwise it is discarded. This same operation is also applied to NNC and DBSCAN.

\subsubsection{Validation of the iDBSCAN parameters}\label{idbscan_vali}

\begin{figure}[ht]
\centering
\includegraphics[width=0.45\textwidth]{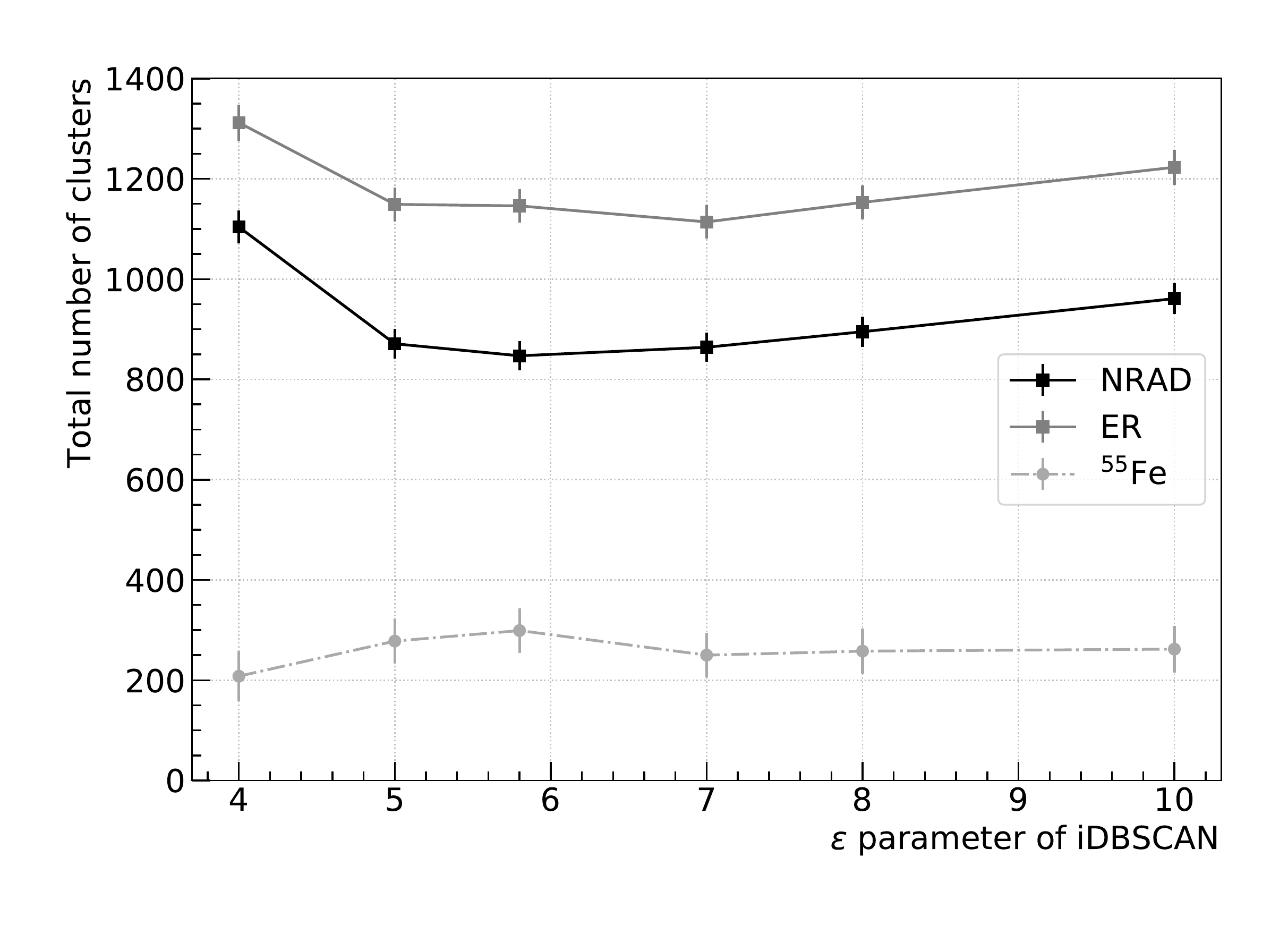}
\includegraphics[width=0.45\textwidth]{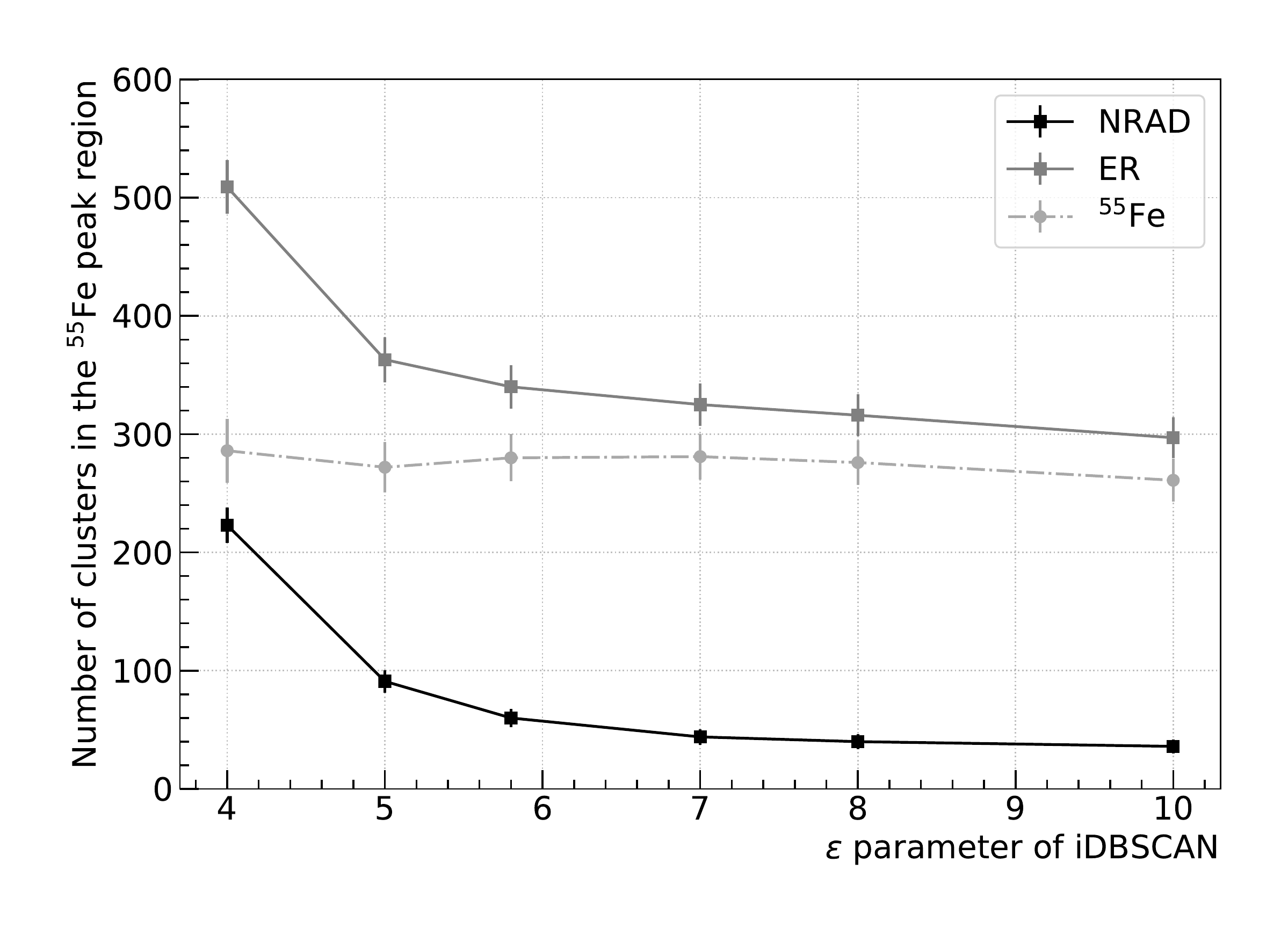}
\caption{Total number of reconstructed clusters (left) and Number of clusters in the $^{55}$Fe peak region (right) as a function of $\epsilon$ for ER and NRAD runs and also a line for the $^{55}$Fe, which means ER-NRAD.}
\label{fig:epsscan}
\end{figure}

The CYGNO Collaboration is currently using iDBSCAN for the clustering method in its event-reconstruction.
The iDBSCAN performance for signals produced by the interactions of photons from $^{55}$Fe has been studied as a function of different values of its parameters: $\epsilon$ and $N_{min}$.
In order to evaluate those values, a test on the detector efficiency and background rejection was carried out: a scan over the two iDBSCAN parameters has been performed.
While the $\epsilon$ ($N_{min}$) parameter will be fixed to a value of 5.8 (30), the other parameter's value will be swept from 5 to 50 (4 to 10). 
Figure \ref{fig:epsscan} (left) shows the total number of clusters found as a function of $\epsilon$ for two distinct datasets: ER and NRAD.
For low $\epsilon$ values the number of NRAD clusters tends to increase, indicating an increase of background contamination. However, for $\epsilon$ values between 5 and 7, this contamination rate stabilizes around a minimum value.
Figure~\ref{fig:epsscan} (right) shows the same trend, while counting only clusters with an integral in the range 2000–4000 photons, characteristic of $^{55}$Fe deposits. This region refers to the energy region of the $^{55}$Fe produced electron recoils (see Fig.~\ref{fig_compfe}).

\begin{figure}[ht]
\centering
\includegraphics[width=0.45\textwidth]{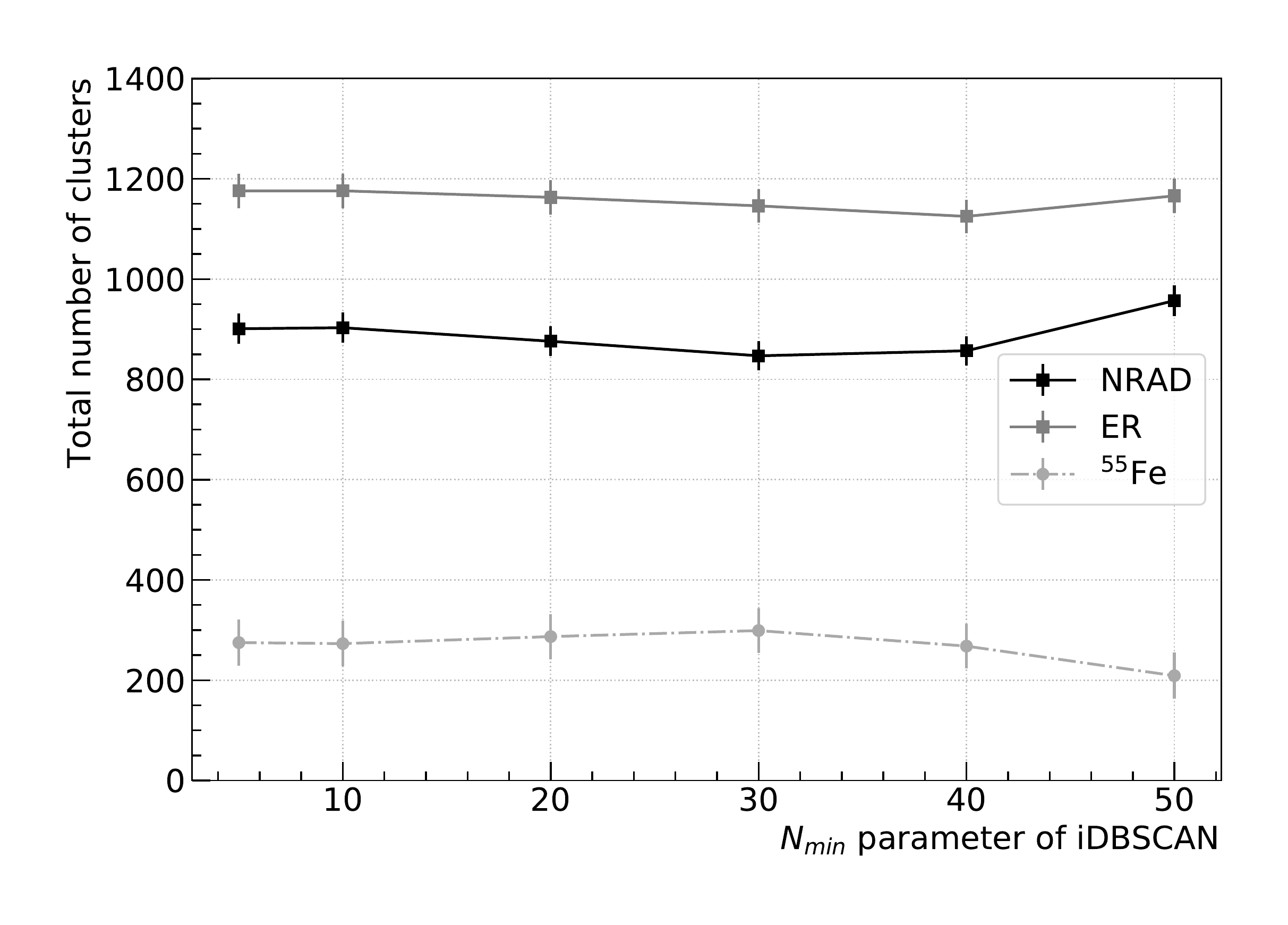}
\includegraphics[width=0.45\textwidth]{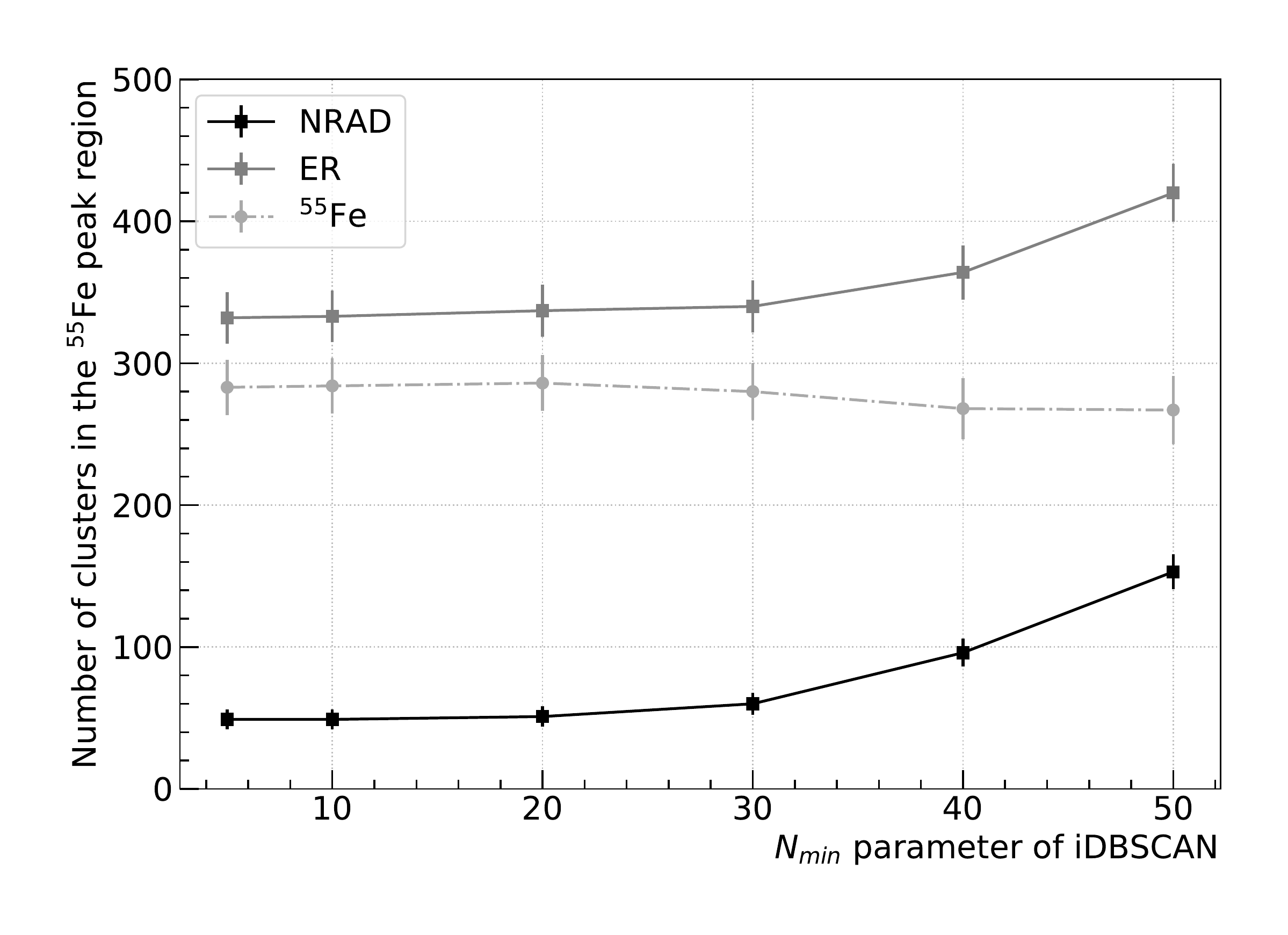}
\caption{Total number of reconstructed clusters (left) and Number of clusters in the $^{55}$Fe peak region (right) as a function of $N_{min}$ for ER and NRAD runs and also a line for the $^{55}$Fe, which means ER-NRAD.}
\label{fig:minscan}
\end{figure}

Similarly, a scan over the $N_{min}$ parameter has been performed as shown in Fig.~\ref{fig:minscan}. Applying the same logic as for the $\epsilon$ parameter, the plot on the left indicates a low contamination region for $N_{min}$ values between 20 and 40, and the right plot to a region for $N_{min}$ $\leq$ 30. In both cases, when stable, the difference between the results indicate a number of $\rm ^{55}Fe$ clusters of about 280.

Finally, energy resolution has also been measured as function of the iDBSCAN parameters.
Values around 12.2\% have been measured for all the $\epsilon$ and $N_{min}$ considered values, with negligible variation. Section \ref{subsec:detres} provides details about the energy resolution measurement. 

\hspace{0pt}

The same procedure performed to choose iDBSCAN parameters was also applied to DBSCAN. The resulting values for the DBSCAN parameters were 6 for $\epsilon$ and 20 for $N_{min}$. It is noteworthy that the value of $\epsilon$ for DBSCAN is very close to the 5.8 found by iDBSCAN, which shows a coherence since the two-dimensional space is the same for both algorithms. The DBSCAN graphs are not shown here as it adds little information to the work considering that they have characteristics similar to those presented in Figs. \ref{fig:epsscan} and \ref{fig:minscan}.

\section{iDBSCAN compared to DBSCAN and NNC}
\label{sec:algoComp}

\subsection{Electronic noise, natural radioactivity and $\rm ^{55}Fe$ energy spectra}

\begin{figure}[ht]
\centering
\includegraphics[width=0.45\textwidth]{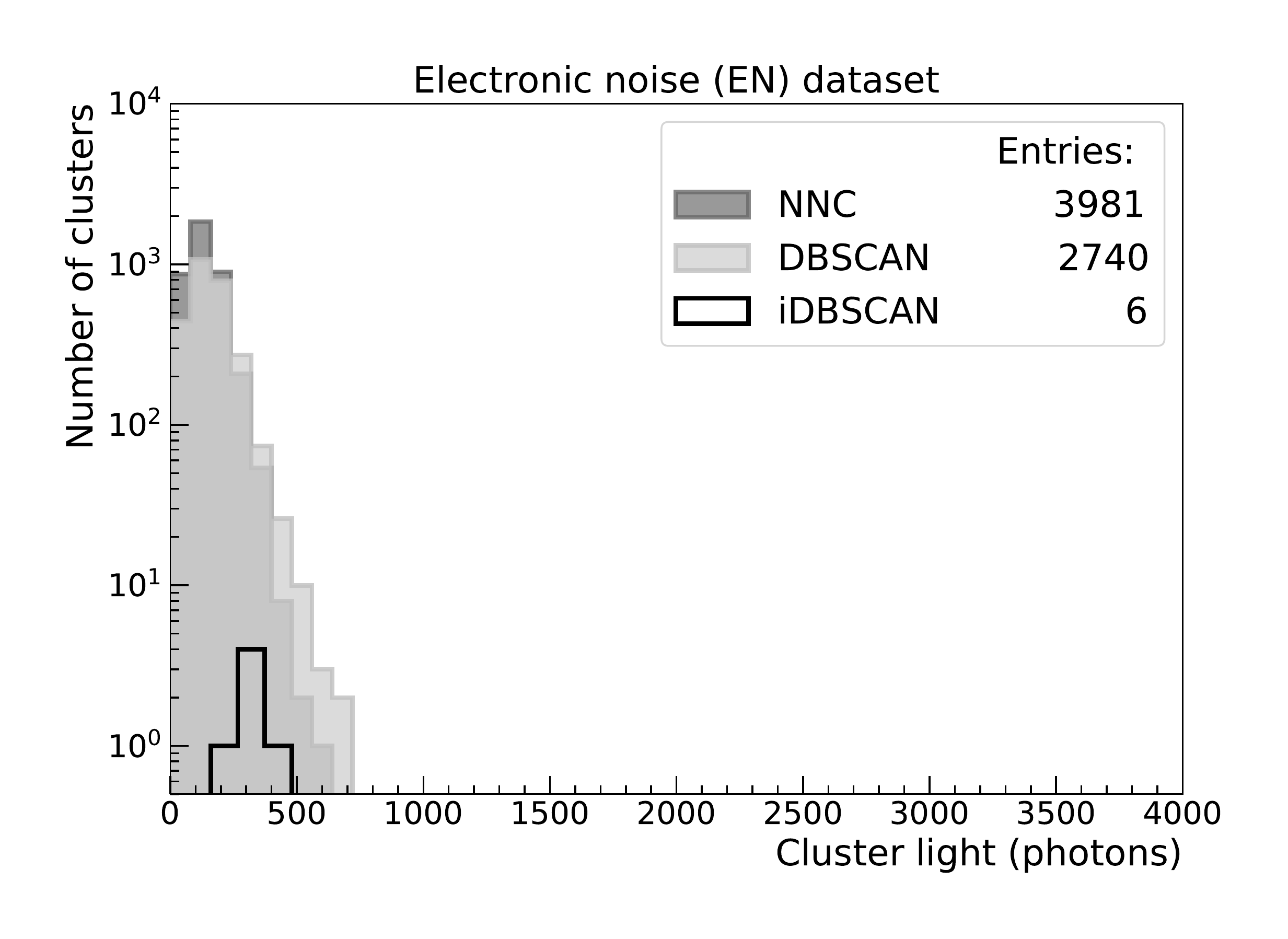}
\includegraphics[width=0.45\textwidth]{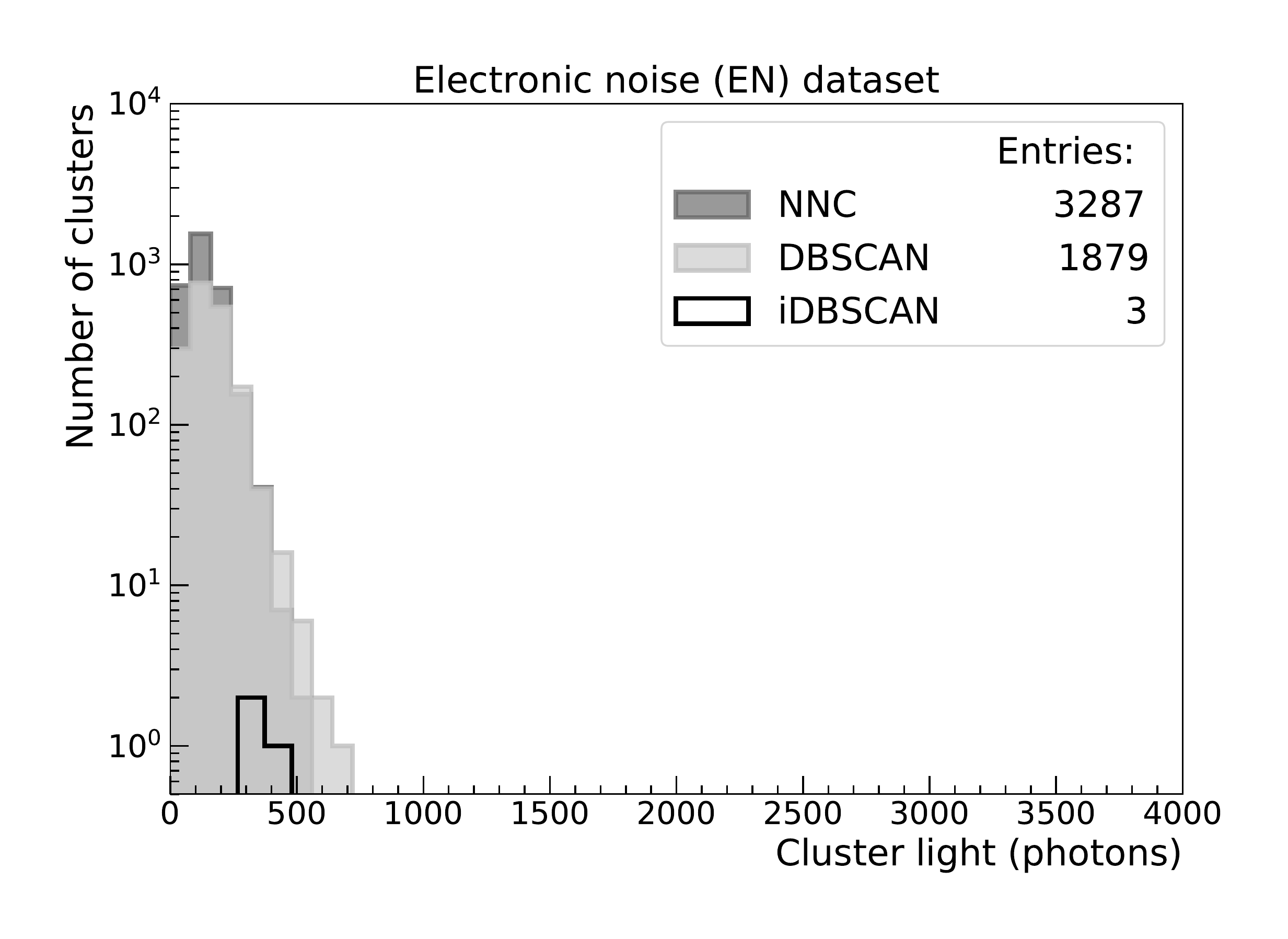}
\caption{Clusters energy distribution for NNC, DBSCAN and iDSBSCAN applied to the EN dataset, without (left) and with (right) a selection on the slimness.}
\label{fig_compnoise}
\end{figure}

\begin{figure}[ht]
\centering
\includegraphics[width=0.45\textwidth]{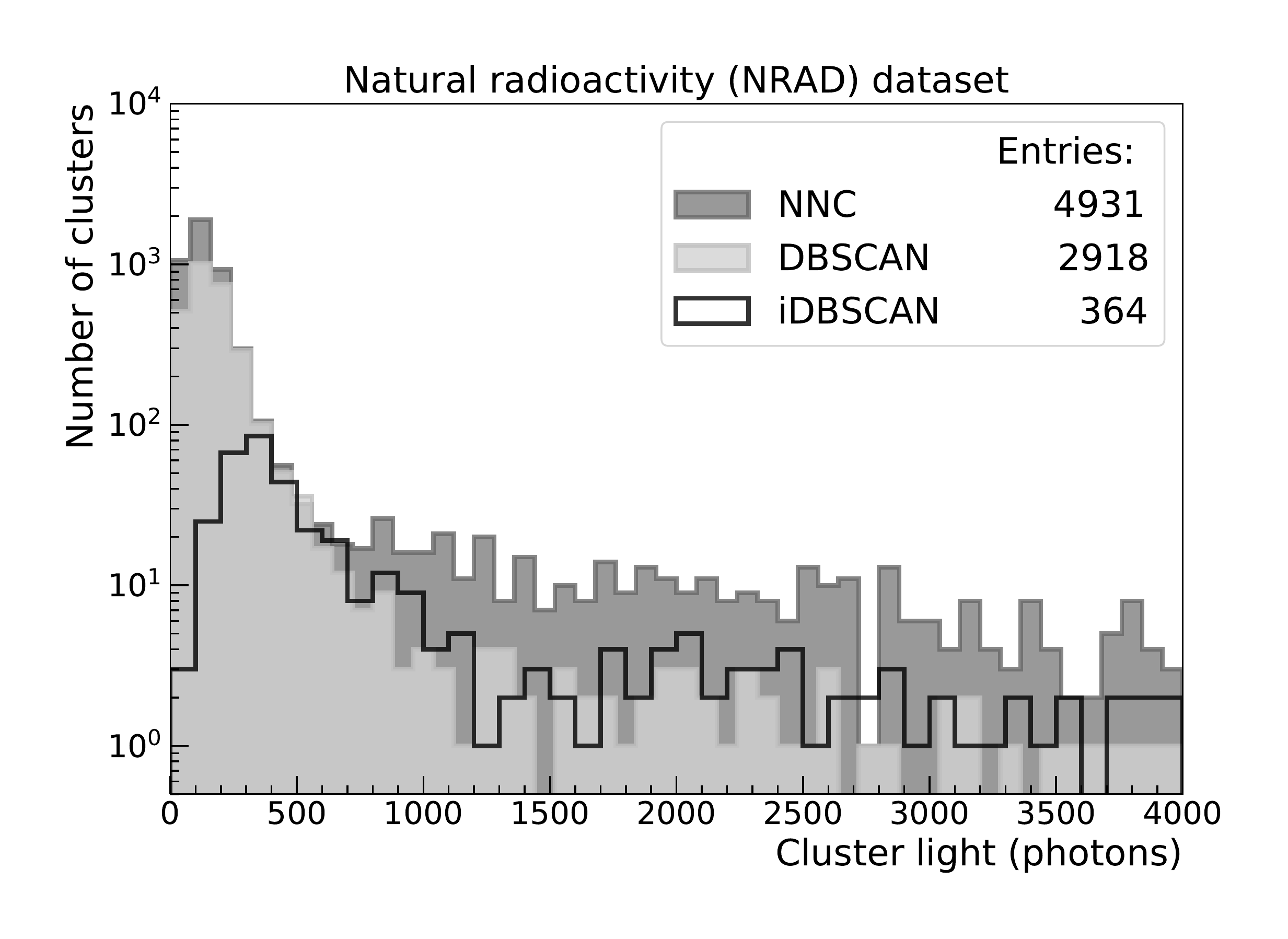}
\includegraphics[width=0.45\textwidth]{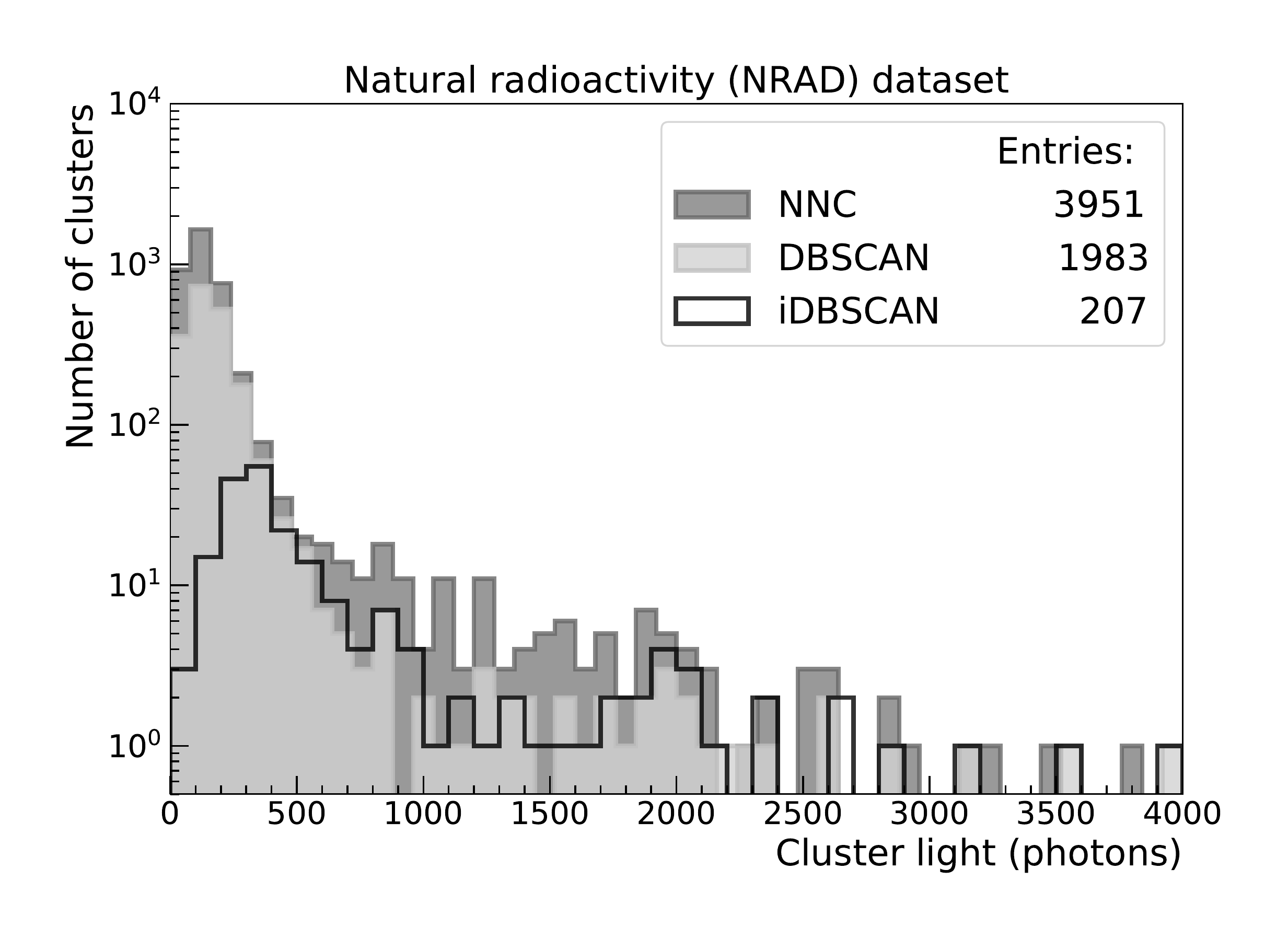}
\caption{Clusters energy distribution for NNC, DBSCAN and iDSBSCAN applied to the NRAD dataset, without (left) and with (right) a selection based on the slimness.}
\label{fig_compcosmic}
\end{figure}

\begin{figure}[ht]
\centering
\includegraphics[width=0.45\textwidth]{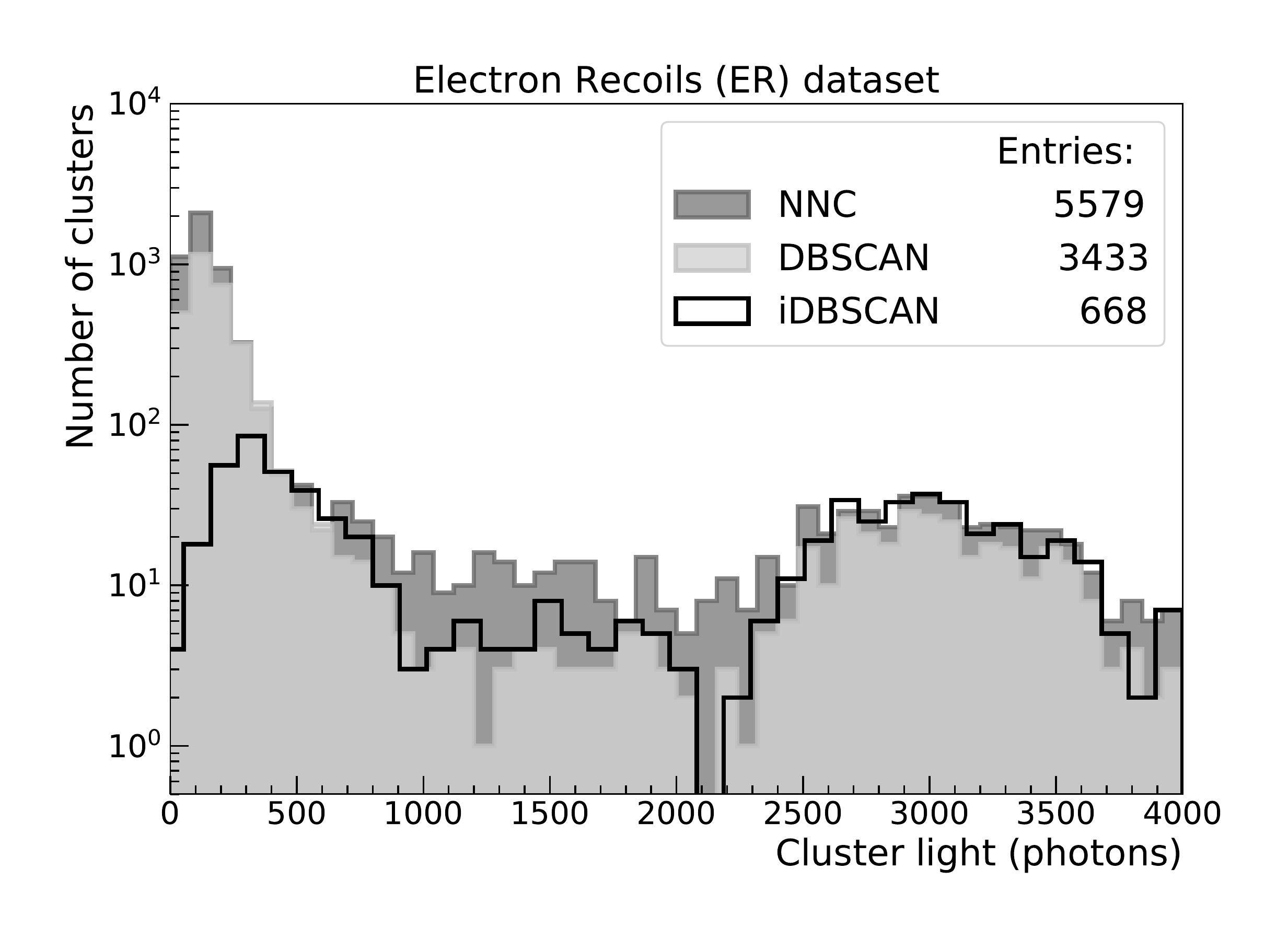}
\includegraphics[width=0.45\textwidth]{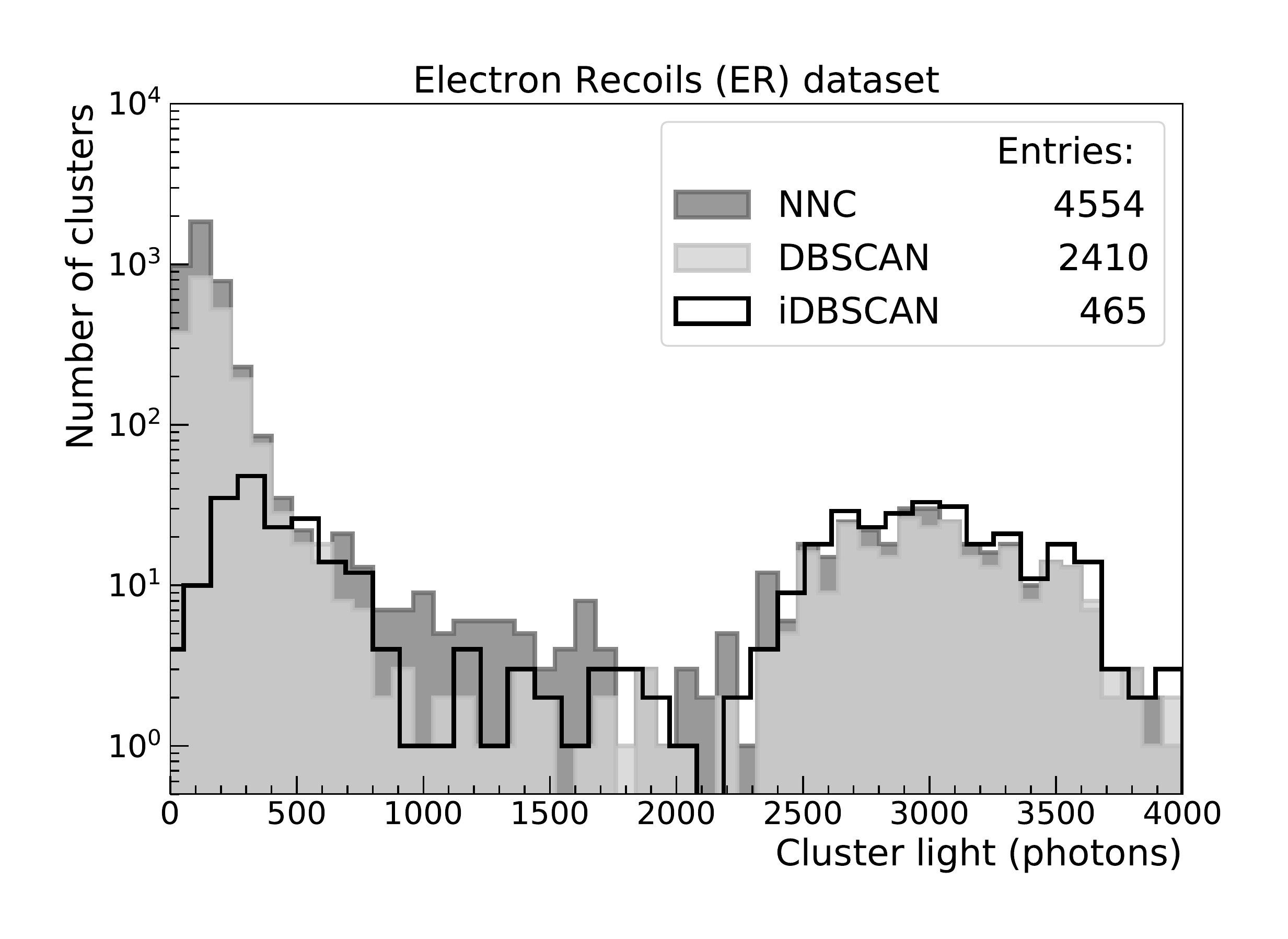}
\caption{Clusters energy distribution for NNC, DBSCAN and iDSBSCAN applied to the ER dataset, without (left) and with (right) a selection based on the slimness.}
\label{fig_compfe}
\end{figure}

The well-known energy deposition signature of 5.9 keV photons coming out from the $\rm ^{55}Fe$ source is exploited in order to evaluate the detection efficiency and background rejection of both methods. While the ER dataset will be used for signal characterization, EN and NRAD datasets will be deployed for background rejection measurements.
The EN acquired data produces low energy clusters with a distribution squeezed in the region below 500 photons as shown in Fig. \ref{fig_compnoise}, NRAD produces an energy distribution widely spread by a heavy tail component as shown in Fig.~\ref{fig_compcosmic} while ER forms an additional narrow distribution centered at around 3000 photons as shown in Fig.~\ref{fig_compfe}.
In this last case, the energy spectrum is composed of background and $^{55}$Fe induced deposits and, therefore, to reconstruct the $^{55}$Fe energy distribution, the background distribution should be subtracted. All the distributions were generated with the same amount of images, 864 of them, except for the iDBSCAN distributions of Fig. \ref{fig_compnoise} which used 6478 images, in order to collect enough EN-clusters, which occur at a low rate.  
Additionally, the signal purity is enhanced accounting for the cluster aspect ratio, called slimness, defined as the ratio between the minor axis (width) and major axis (length) of each cluster.

Figure \ref{fig_compnoise} compares the energy spectrum of clusters generated by NNC and DBSCAN with those generated by iDBSCAN for EN events without and with a selection based on the slimness parameter, considering only clusters with slimness greater than 0.4 for the latter case. The computed numbers of EN-clusters per image for NNC, DBSCAN and iDBSCAN were $4.61 \pm 0.17$, $3.17 \pm 0.12$ and $(9 \pm 4)\times10^{-4}$, respectively.
Regarding NNC and DBSCAN, EN-clusters dominate the background rate for energies below 500 photons which can be noticed by comparing the EN energy distribution of Fig. \ref{fig_compnoise} with that of the NRAD shown in Fig. \ref{fig_compcosmic}. Selection on slimness variable decreases the number of clusters per image to $3.80 \pm 0.14$, $2.17 \pm 0.09$ and $(5 \pm 3)\times 10^{-4}$ for NNC, DBSCAN and iDBSCAN, respectively. Therefore, when compared to NNC and DBSCAN, iDBSCAN is able to reduce the number of EN-clusters per image by a factor of $(3\div7)\times 10^3$.

Figure~\ref{fig_compcosmic} shows the energy distributions for the NNC, DBSCAN and iDBSCAN clusters using the NRAD dataset without (left) and with (right) a selection on slimness.
iDBSCAN presents a clear peak evolution around 300 photons while NNC and DBSCAN accumulate clusters with lower energies due to EN-clusters.
iDBSCAN and DBSCAN reduce the number of background events in the region between 2000 and 4000 photons when compared to NNC, which is the region where the $\rm ^{55}Fe$ events are expected to be, as mentioned before, providing better background rejection for low energy events as for the 5.9 keV photons.
On the right of Fig.~\ref{fig_compcosmic}, the distribution of light, only considering clusters with slimness greater than 0.4 is shown. This selection reduces even more the number of background events in the $\rm ^{55}Fe$ region, bringing NNC closer to the other methods. 
However, for the lower energy region, the number of fake clusters is only slightly reduced, causing iDBSCAN to maintain a better background rejection efficiency when compared to NNC and DBSCAN.

Figure~\ref{fig_compfe} shows the results of the same analysis performed on the ER dataset. In this case, the  sum of the distribution obtained in the NRAD sample and the one from $^{55}$Fe interactions is expected.
As shown, all three clustering algorithms are sensitive to the 5.9 keV photon events. However, as commented previously, a higher purity level is achieved using iDBSCAN.
After applying the slimness threshold, as shown in the right plot of Fig.~\ref{fig_compfe}, the distributions around the $^{55}$Fe peak of NNC, DBSCAN and iDBSCAN get closer indicating that the three methods have similar detection efficiency considering that the number of $^{55}$Fe spots found by each method is practically the same.  

\subsection{Slimness selection optimization}
\begin{figure}[ht]
\centering
\includegraphics[width=0.32\textwidth]{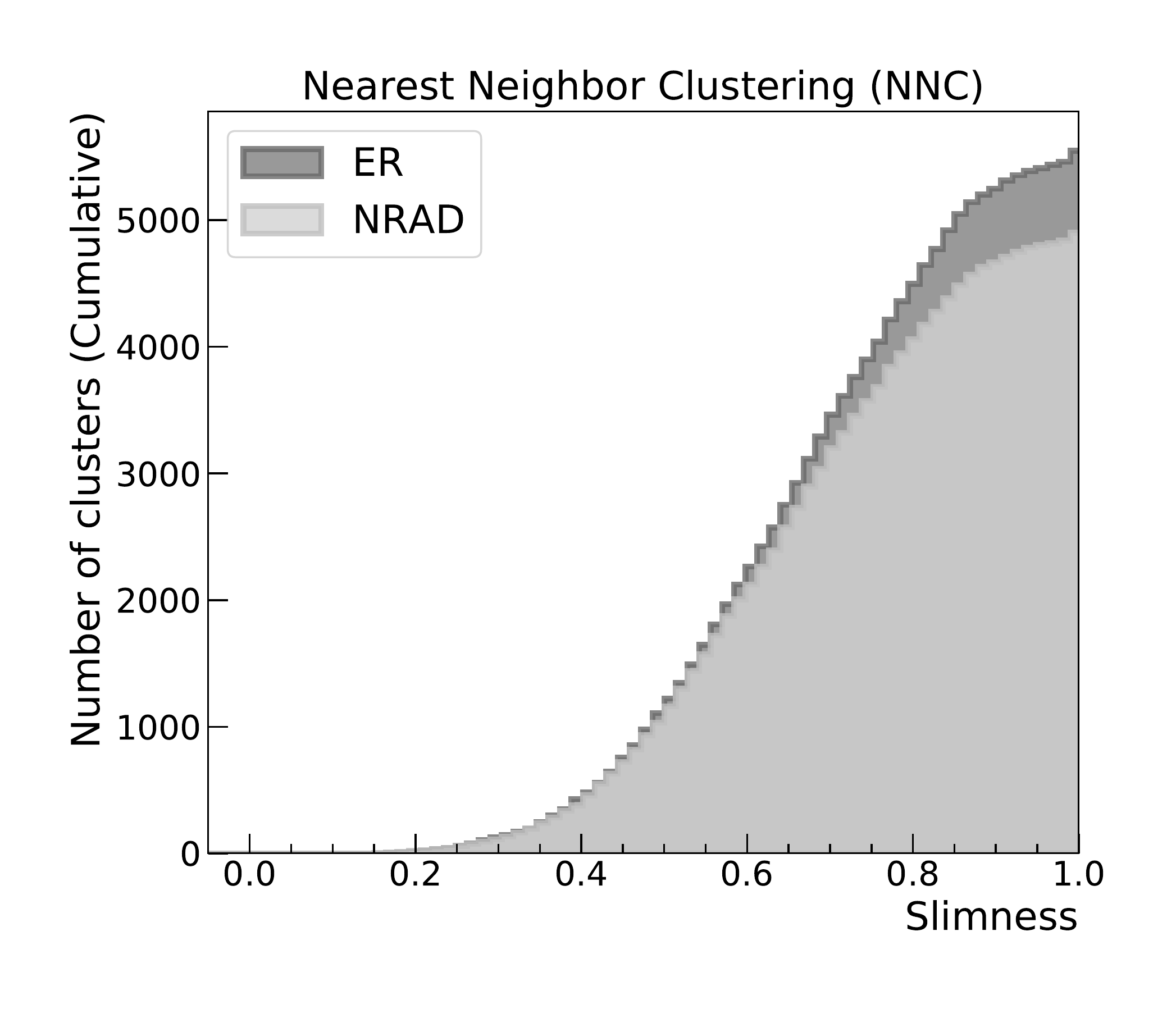}
\includegraphics[width=0.32\textwidth]{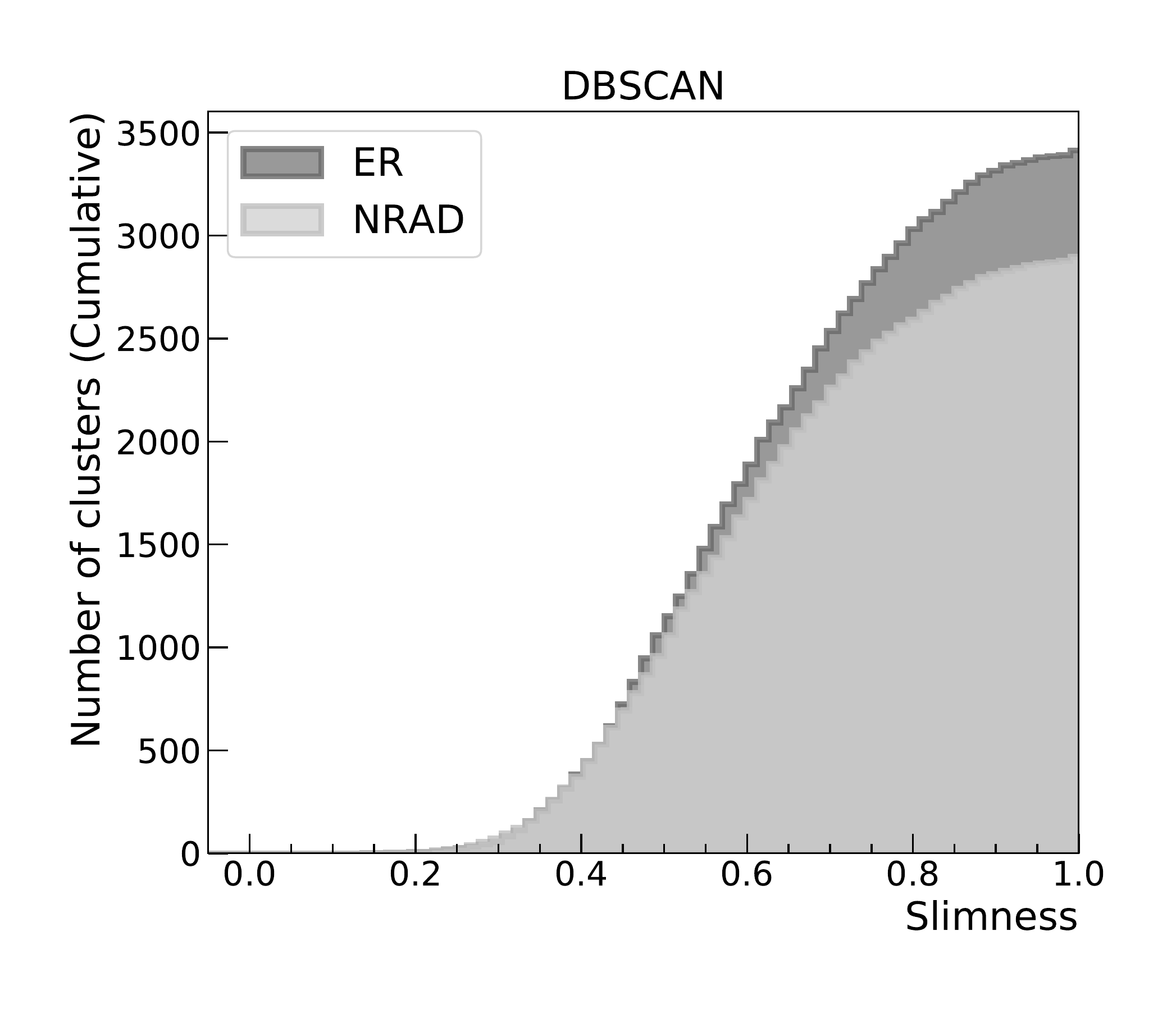}
\includegraphics[width=0.32\textwidth]{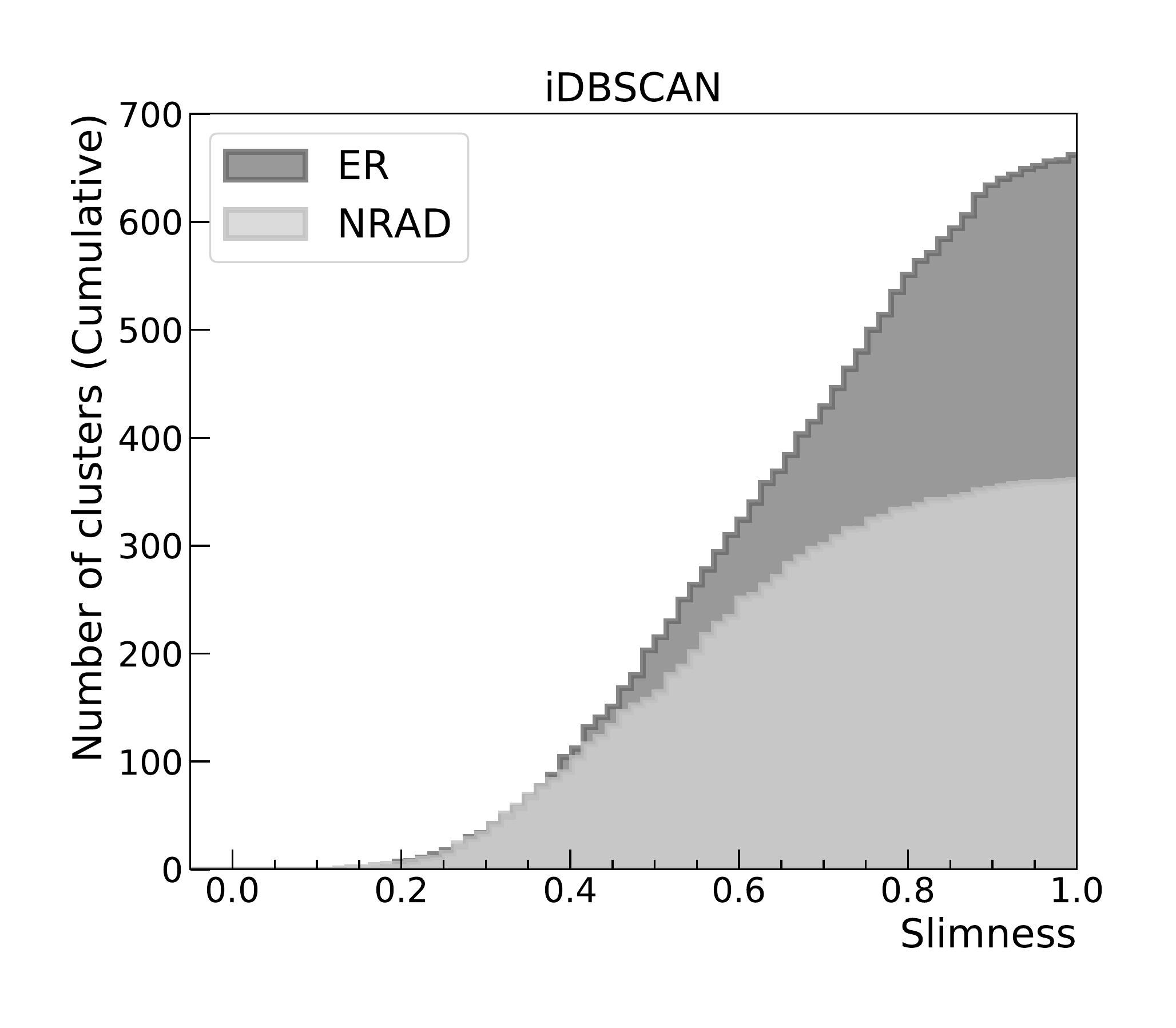}
\caption{Cumulative distribution of the slimness for NRAD and ER data, for NNC, DBSCAN and iDSBSCAN.} 
\label{fig_cdf_slim}
\end{figure}

Figure \ref{fig_cdf_slim} shows the slimness cumulative distribution of clusters for an interval between 0 and 1, applied to the NRAD and ER datasets for NNC, DBSCAN and iDBSCAN.
As it is possible to see, in all cases $^{55}$Fe spots tend to have slimness higher than about 0.4.
This variable can be used in conjunction with energy measurement to discriminate $^{55}$Fe spots from background clusters.
In this section the value of slimness will be swept so that it is possible to choose the most suitable value for its use as an event selection parameter as well as to evaluate its impact when applied together with the energy measurement.

\begin{figure}[ht]
\centering
\includegraphics[width=0.32\textwidth]{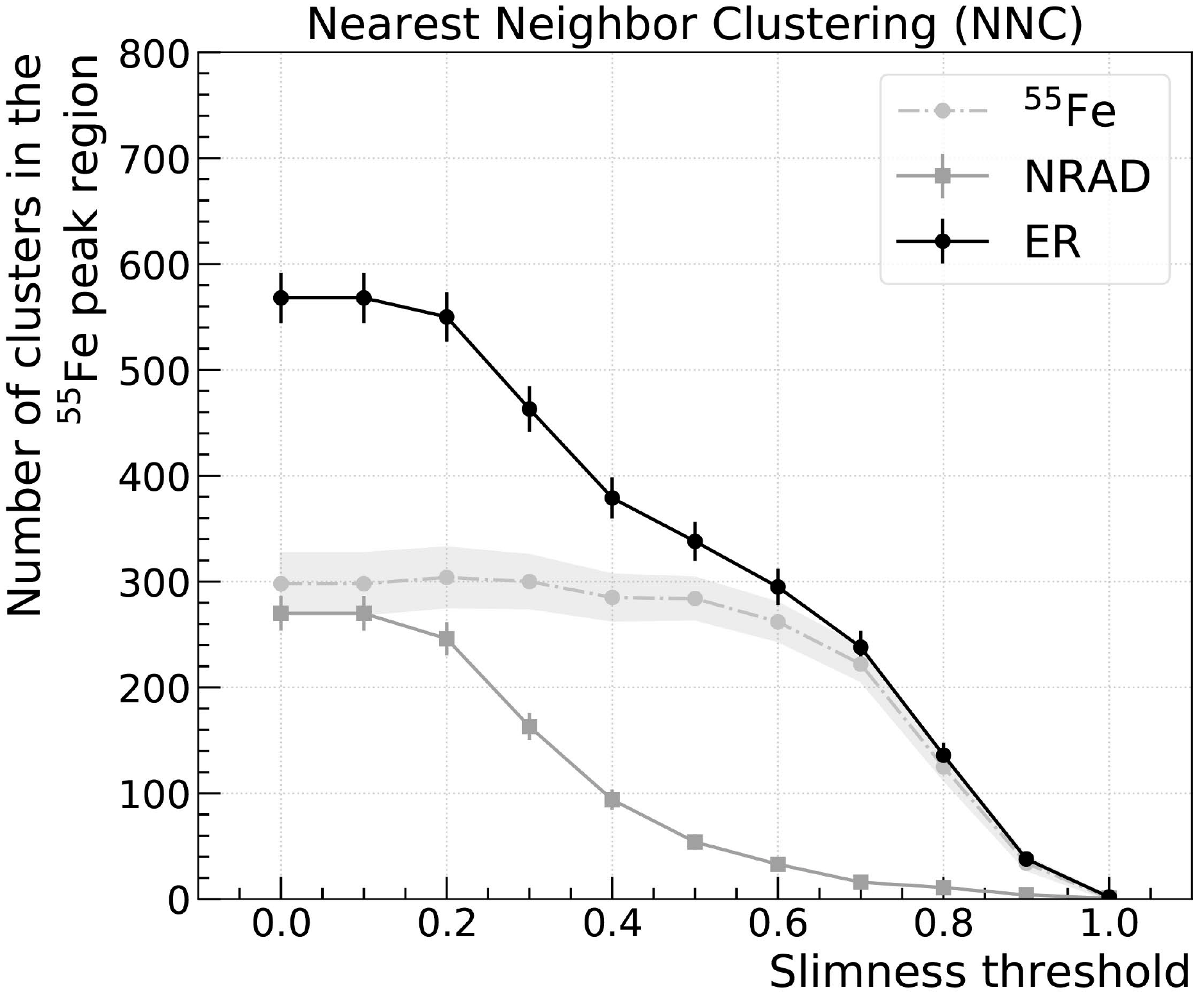}
\includegraphics[width=0.32\textwidth]{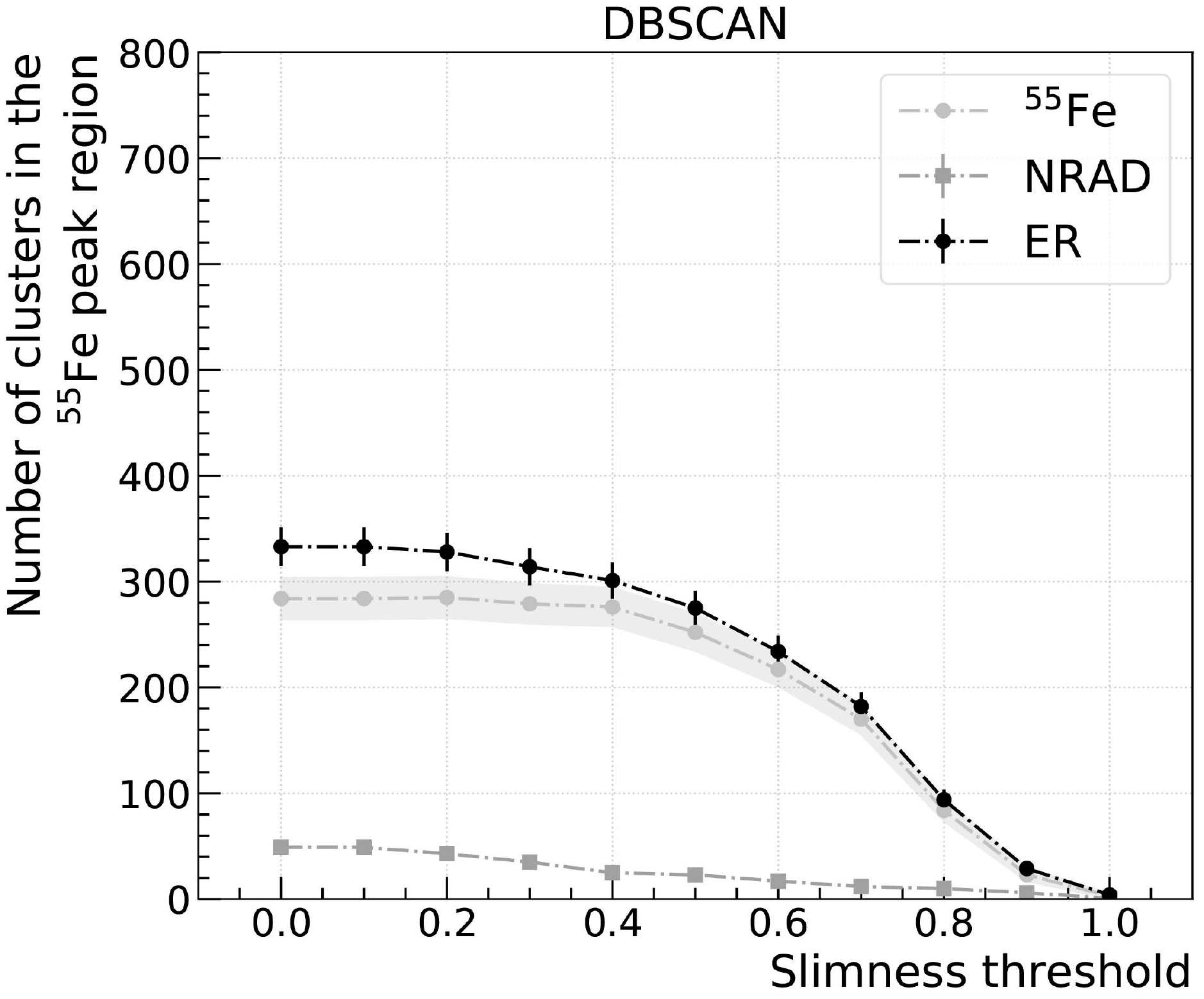}
\includegraphics[width=0.32\textwidth]{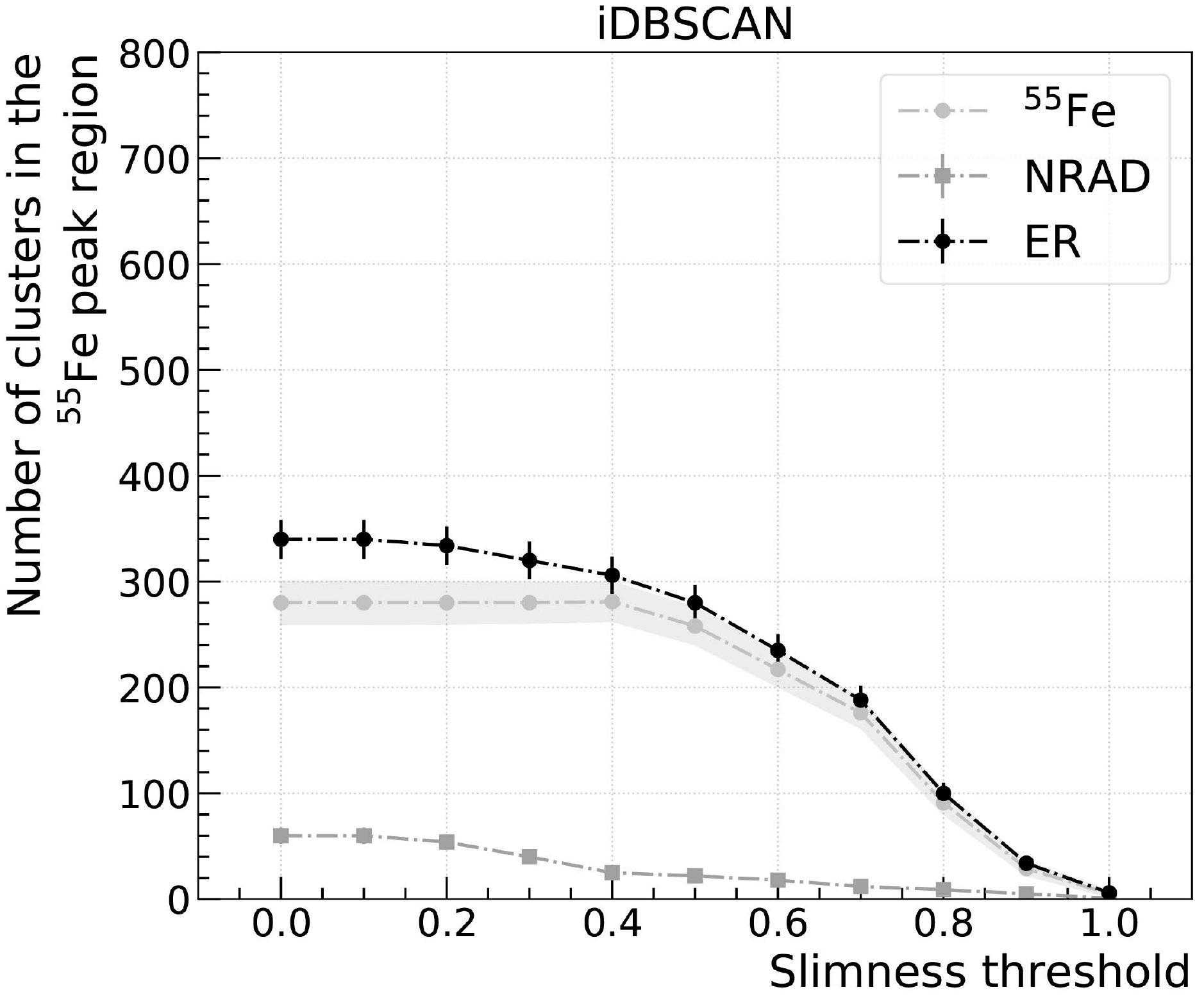}
\caption{Scan in the number of clusters on the $\rm ^{55}Fe$ peak region (between 1500 and 4500 photons) when changing the threshold on the slimness for NRAD and ER data, for NNC, DBSCAN and iDSBSCAN.} 
\label{fig_slim_scan}
\end{figure}

In order to evaluate the signal efficiency and purity as a function of the slimness selection for the two algorithms, the number of clusters within the selected $^{55}$Fe energy region (from 1500 to 4500 photons) was measured for various slimness threshold values (X $\geqslant$ x) as shown in Fig.~\ref{fig_slim_scan} for the NNC, DBSCAN and iDBSCAN algorithms.
This figure shows that DBSCAN and iDBSCAN find a similar number of clusters in the $^{55}$Fe region when compared to NNC for slimness below 0.4, given by the difference between the ER and NRAD curves, but with lesser contamination (NRAD curves).

Considering that the $^{55}$Fe clusters produce an intensity that follows a Gaussian distribution with an average value of about 3000 photons and standard deviations of 550, 385 and 371, for NNC, DBSCAN and iDBSCAN respectively (see Fig. \ref{fig_CosFe}), then more than 99\% of the $^{55}$Fe clusters are selected between 1500 and 4500 photons.
On the other hand, for the same region, the subtraction of the natural radioactivity events between the ER and NRAD acquisition runs has a mean value equal to zero but a fluctuation of about 23 (14), 10 (7) and 11 (7) clusters for slimness equal to 0.0 (0.4), for NNC, DBSCAN and iDBSCAN respectively. Therefore, the dashed line of Fig. \ref{fig_slim_scan} is composed mainly of $^{55}$Fe events plus few background events produced by the statistical fluctuation that occurs in the process of subtracting natural radioactivity.
As can be noticed by observing Fig.~\ref{fig_compcosmic}, DBSCAN and iDBSCAN tend to have less background contamination than NNC, reducing the statistical uncertainty related to the background subtraction. This effect is also shown by the shaded band drawn around the dashed lines of Fig. \ref{fig_slim_scan}.

Based on the measurements of Fig.~\ref{fig_slim_scan}, the impact of the slimness parameter can be assessed by measuring the relative efficiency ($\rm \varepsilon_{sel}$) with respect to the bin with the highest content in the $^{55}$Fe curve (so that for such a bin, $\rm \varepsilon_{sel}$ = 100\%), and fake events ($\rm F_{evts}$), as defined below:

\begin{itemize}
    \item $\rm \varepsilon_{sel}$: number of clusters found in the ER dataset ($\rm nFe$) subtracted by the number of clusters found in the NRAD dataset ($\rm nRd$) divided by the maximum value of the $\rm nFe-nRd$ subtraction among all slimness values (see Equation~\ref{eq:01});
    
    \begin{equation}
       {\rm \varepsilon_{sel} = \left( {\frac{{nFe - nRd}}{{\max \left( {nFe - nRd} \right)}}} \right)}
       \label{eq:01}
    \end{equation}

    \item $\rm F_{evts}$: ratio between the number of clusters found in the NRAD dataset ($\rm nRd$) and the number of clusters found in the ER dataset ($\rm  nFe$) (see Equation~\ref{eq:02}a). This measure can also be understood in terms of background rejection ($\rm B_{rj}$) as shown by Equation~\ref{eq:02}b;
    
\end{itemize}
\begin{equation}
    \begin{array}{*{20}{c}}
   \begin{array}{*{20}{c}}
   {\rm F_{evts} = \left( {\frac{{nRd}}{{nFe}}} \right)}  \\
\end{array} ~~(a) & {, ~~~} & {\rm B_{rj} = 1 - F_{evts} }  \\
\end{array} ~~(b)
\label{eq:02}
\end{equation}

Figure~\ref{fig_cdf_slim} shows that for slimness below 0.4 the efficiency for background events is very small, while most of the $^{55}$Fe events are retained.
Tables~\ref{tab:effComp} and ~\ref{tab:falseComp} shows, respectively, the computed $\rm \varepsilon_{sel}$ and $\rm F_{evts}$ for both clustering methods and different thresholds on the slimness variable ranging from 0.0 to 0.8. The errors presented in these tables were computed considering a confidence interval of 95\% for a binomial proportion \cite{brown2001interval}. For the high efficiency region ($\geq 0.94$), occurring for slimness values from 0.0 to 0.4, iDBSCAN and DBSCAN achieved a lower fake event probability, always about 3 times less than NNC. For slimness greater than or equal to 0.6 all methods begin to lose efficiency.
More specifically, for a slimness threshold of 0.4, the efficiency is still close to 100\%, compared to not using slimness, but the number of fake events is reduced by a factor of about 2 for all the methods.

\begin{table}[ht]
\small
\centering
\caption{$\rm \varepsilon_{sel}$ comparison between iDBSCAN, DBSCAN and NNC.}
\label{tab:effComp}
\begin{tabular}{c|cc|cc|cc}
\multirow{2}{*}{\begin{tabular}[c]{@{}c@{}}\textbf{Slimness}\\ (width/length)\end{tabular}} & \multicolumn{6}{c}{$\mathbf{\rm \varepsilon_{sel}}$}                  \\
                                                                                       & \multicolumn{2}{c|}{iDBSCAN} & \multicolumn{2}{c|}{DBSCAN}   & \multicolumn{2}{c}{NNC} \\ \hline \hline
0.0 & 1.00  & $^{+0.00}_{-0.02}$ & 1.00 & $^{+0.00}_{-0.02}$ & 0.98  & $^{+0.01}_{-0.02}$\\
0.2 & 1.00  & $^{+0.00}_{-0.02}$ & 1.00 & $^{+0.00}_{-0.01}$ & 1.00  & $^{+0.00}_{-0.01}$\\
0.4 & 1.00  & $^{+0.00}_{-0.01}$ & 0.97 & $^{+0.02}_{-0.03}$ & 0.94  & $^{+0.02}_{-0.03}$\\
0.6 & 0.77  & $^{+0.05}_{-0.05}$ & 0.76 & $^{+0.05}_{-0.05}$ & 0.86  & $^{+0.03}_{-0.04}$\\
0.8 & 0.32  & $^{+0.06}_{-0.05}$ & 0.29 & $^{+0.05}_{-0.05}$ & 0.41  & $^{+0.05}_{-0.06}$
\end{tabular}
\end{table}

\begin{table}[ht]
\small
\centering
\caption{$\rm F_{evts}$ comparison between iDBSCAN, DBSCAN and NNC.}

\label{tab:falseComp}
\begin{tabular}{c|cc|cc|cc|c|c}

\multirow{2}{*}{\begin{tabular}[c]{@{}c@{}}\textbf{Slimness}\\ (width/length)\end{tabular}} & \multicolumn{6}{c|}{$F_{evts}$}  & \multicolumn{2}{c}{\begin{tabular}[c]{@{}c@{}} iDBSCAN $B_{rj}$ variation (\%) \end{tabular}}                  \\
& \multicolumn{2}{c|}{iDBSCAN} & \multicolumn{2}{c|}{DBSCAN}   & \multicolumn{2}{c|}{NNC} & \multicolumn{1}{c|}{~~~~DBSCAN~~~~~} & \multicolumn{1}{c}{NNC}   \\ \hline \hline
0.0 & 0.18  & $^{+0.04}_{-0.04}$ & 0.15 & $^{+0.04}_{-0.04}$ & 0.48 & $^{+0.04}_{-0.04}$  & ~-3.4  $^{~~~~+7.1}_{~~~~-6.5}$ & ~57.0 $^{~~~+11.5}_{~~~-12.3}$\\

0.2 & 0.16  & $^{+0.04}_{-0.04}$ & 0.13 & $^{+0.04}_{-0.03}$ & 0.45 & $^{+0.04}_{-0.04}$  & ~-3.5 $^{~~~~+6.9}_{~~~~-6.4}$ & ~51.7 $^{~~~+10.7}_{~~~-11.6}$\\

0.4 & 0.08  & $^{+0.04}_{-0.03}$ & 0.08 & $^{+0.04}_{-0.03}$ & 0.25 & $^{+0.05}_{-0.04}$  & ~~0.1 $^{~~~~+5.3}_{~~~~-6.4}$ & ~22.6 $^{~~~~+6.6}_{~~~~-8.0}$ \\

0.6 & 0.08  & $^{+0.04}_{-0.03}$ & 0.07 & $^{+0.04}_{-0.03}$ & 0.11 & $^{+0.04}_{-0.03}$ & ~-0.4 $^{~~~~+6.4}_{~~~~-4.7}$ & ~~~4.0 $^{~~~~+4.9}_{~~~~-6.7}$\\

0.8 & 0.09  & $^{+0.07}_{-0.04}$ & 0.11 & $^{+0.08}_{-0.05}$ & 0.08 & $^{+0.06}_{-0.04}$ & ~~1.8 $^{~~~~+6.5}_{~~~-10.5}$ & ~~-1.0 $^{~~~+10.2}_{~~~~-6.4}$

\end{tabular}
\end{table}

The last column of Table \ref{tab:falseComp} shows the iDBSCAN background-rejection improvement compared to NNC.
For slimness equal to 0.4, for example, iDBSCAN has 92\% of background rejection efficiency while NNC has 75\%, leading to a relative improvement of (92-75)/75 $\approx$ 23\%.
Finally, The second-last column of this same table shows that iDBSCAN and DBSCAN present similar background-rejection performances.

\subsection{Light Yield Resolution}\label{subsec:detres}

The detector energy resolution was estimated by a fit to the clusters energy distributions accounting for natural radioactivity and the $^{55}$Fe events. The former was modeled by an exponential function and the latter by a Polya function \cite{bib:rolandiblum}:

\begin{equation}
   P(n)=\frac{1}{b\overline{n}}\frac{1}{k!}\left(\frac{n}{b\overline{n}}\right)^k \cdot e^{-n/b\overline{n}}
\label{fun:polya}
\end{equation}
where $b$ is a free parameter and $k=1/b-1$. The distribution has $\overline{n}$ as expected value, while the variance is governed by $\overline{n}$ and the $b$ parameter, as follows: $\sigma^2=\overline{n}(1+b\overline{n})$. The total likelihood is given by the sum of the two functions.

\begin{figure}[ht]
\centering
\includegraphics[width=0.32\textwidth]{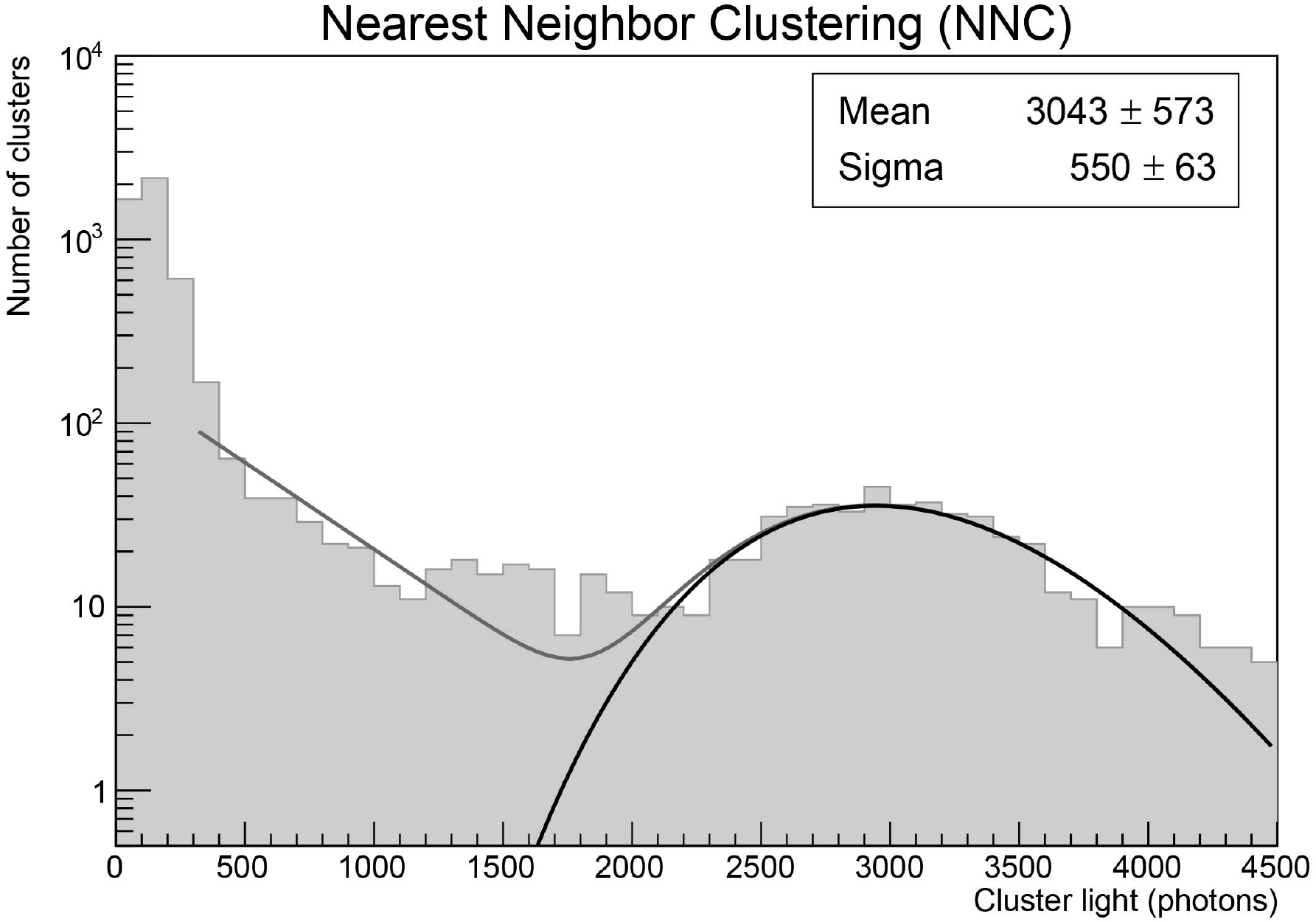}
\includegraphics[width=0.32\textwidth]{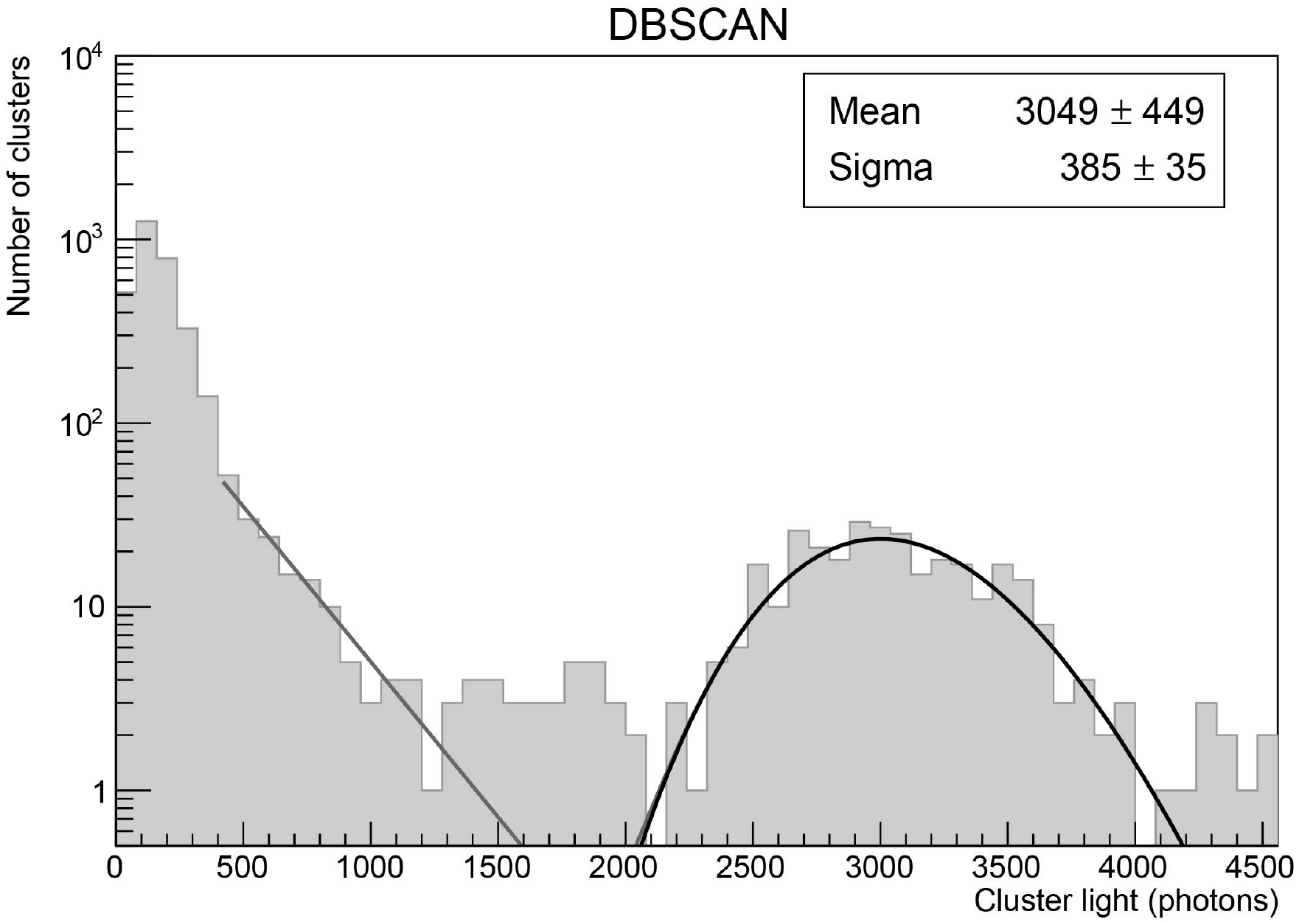}
\includegraphics[width=0.32\textwidth]{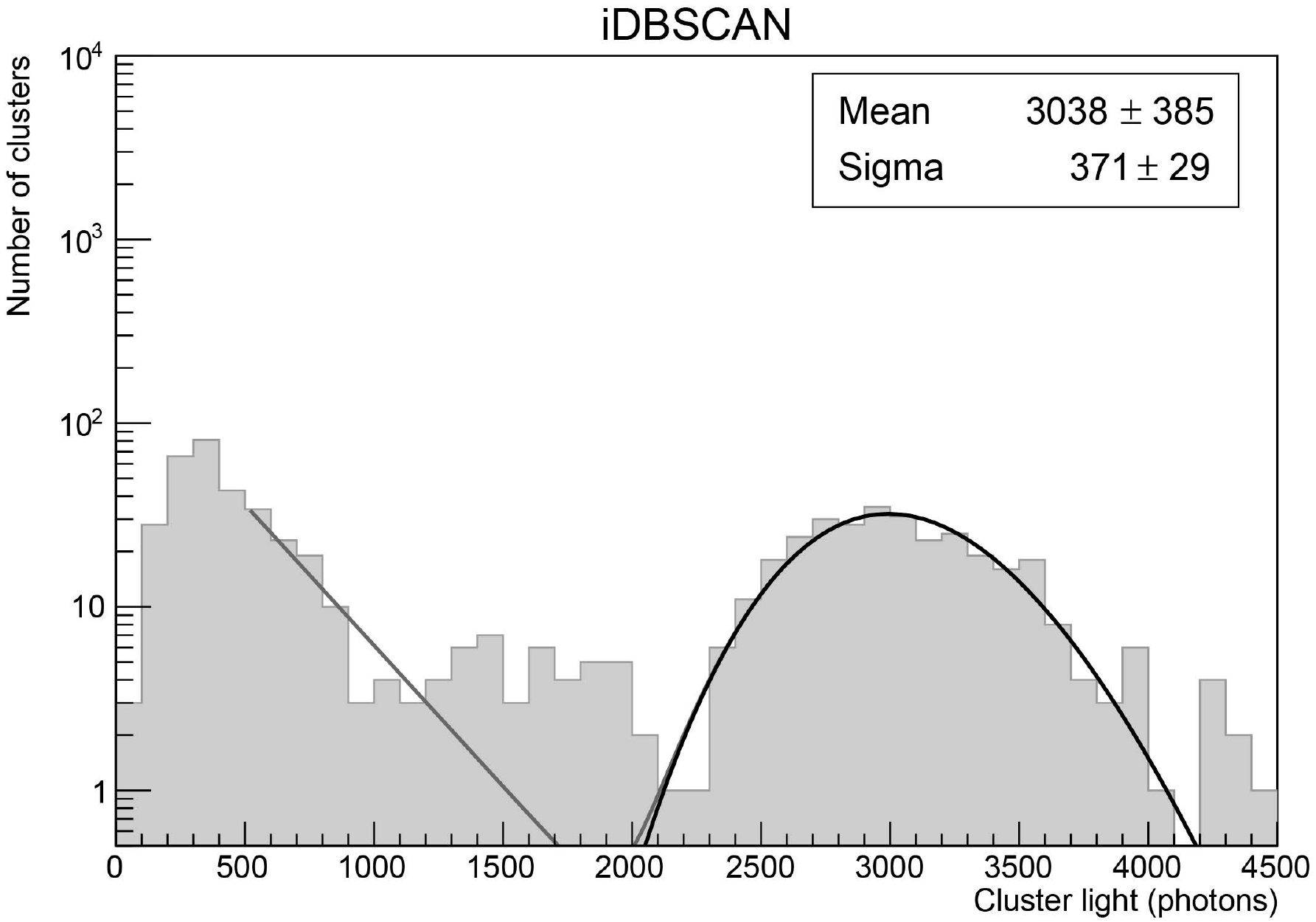}
\caption{Results of the fit applied to the NNC, DBSCAN and iDBSCAN energy distributions.}
\label{fig_CosFe}
\end{figure}

Figure~\ref{fig_CosFe} shows the fit results for NCC, DBSCAN and iDBSCAN clusters without applying any selection on the slimness parameter.
Based on the computed values, energy resolution were measured to be (18.1 $\pm$ 3.9)\%, (12.6 $\pm$ 2.2)\% and (12.2 $\pm$ 1.8)\% for NNC, DBSCAN and iDBSCAN respectively, and the energy conversion factor approximately 515 ADC units per keV for all of them.
Conversion factor and energy resolution are computed using the \textit{mean} and \textit{sigma} parameters shown in Fig.~\ref{fig_CosFe}. 
The former is the \textit{mean} divided by 5.9 keV (ER energy), while the latter is given by dividing the \textit{sigma} by the \textit{mean}.

\begin{figure}[ht]
\centering
\includegraphics[width=0.32\textwidth]{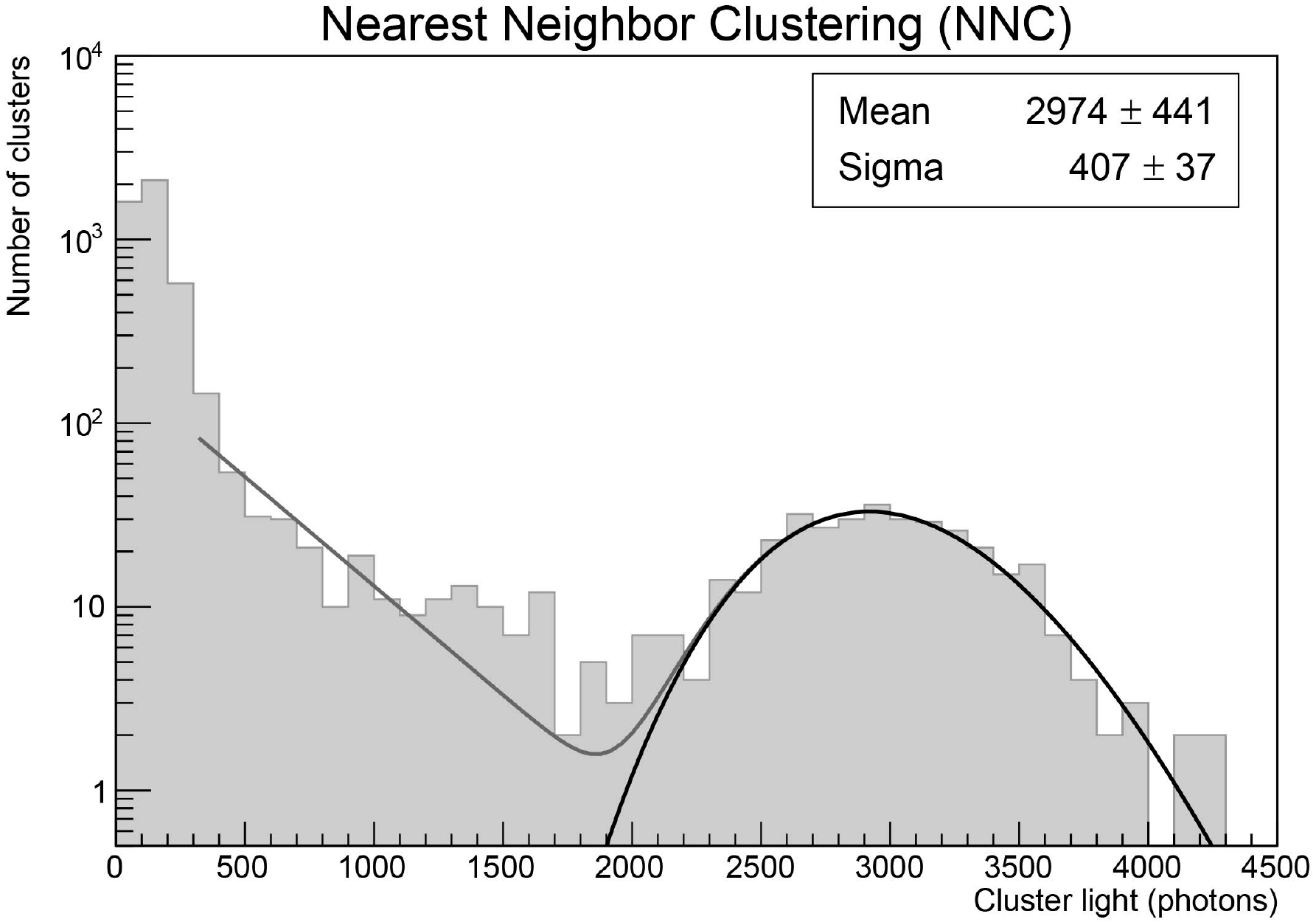}
\includegraphics[width=0.32\textwidth]{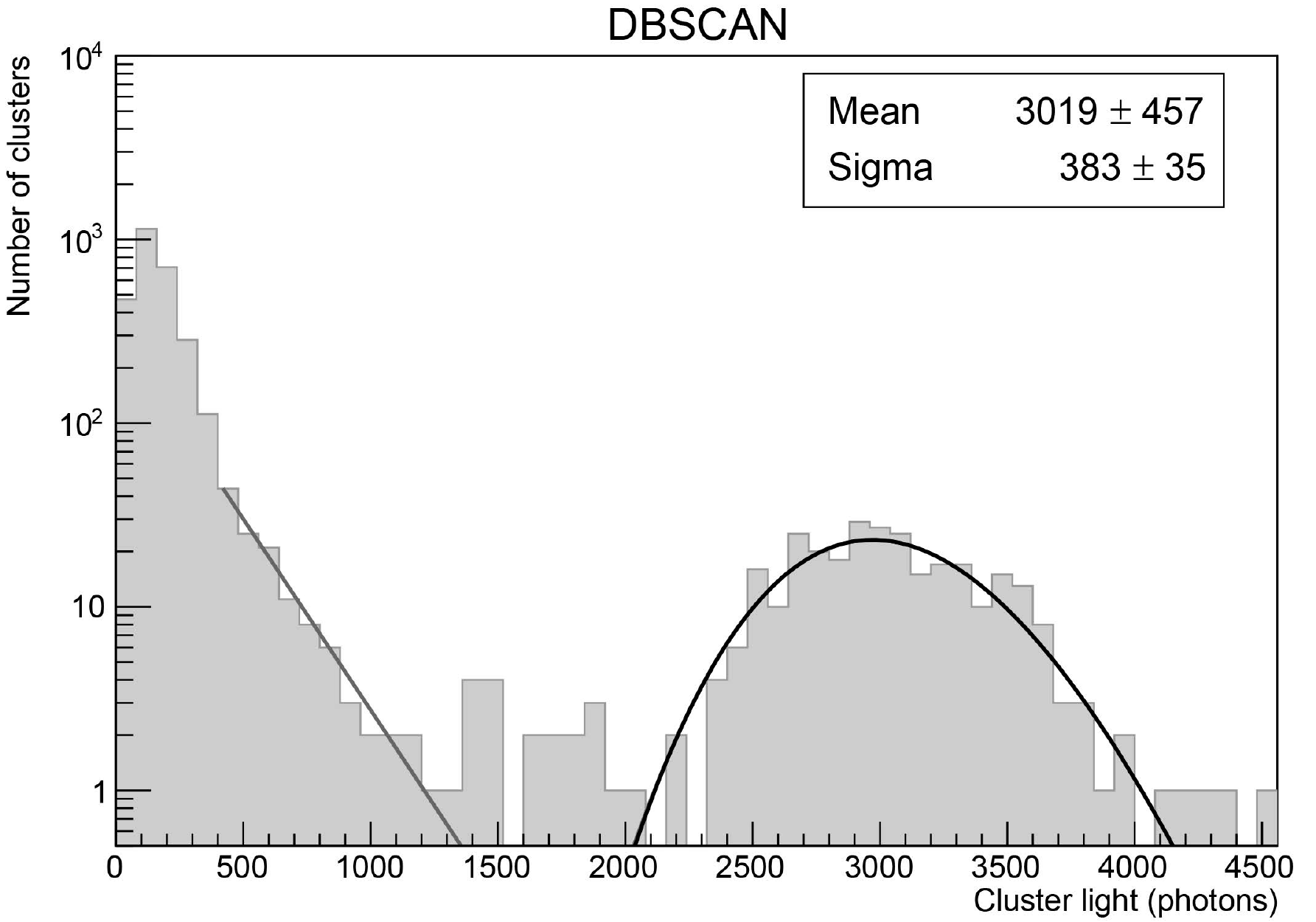}
\includegraphics[width=0.32\textwidth]{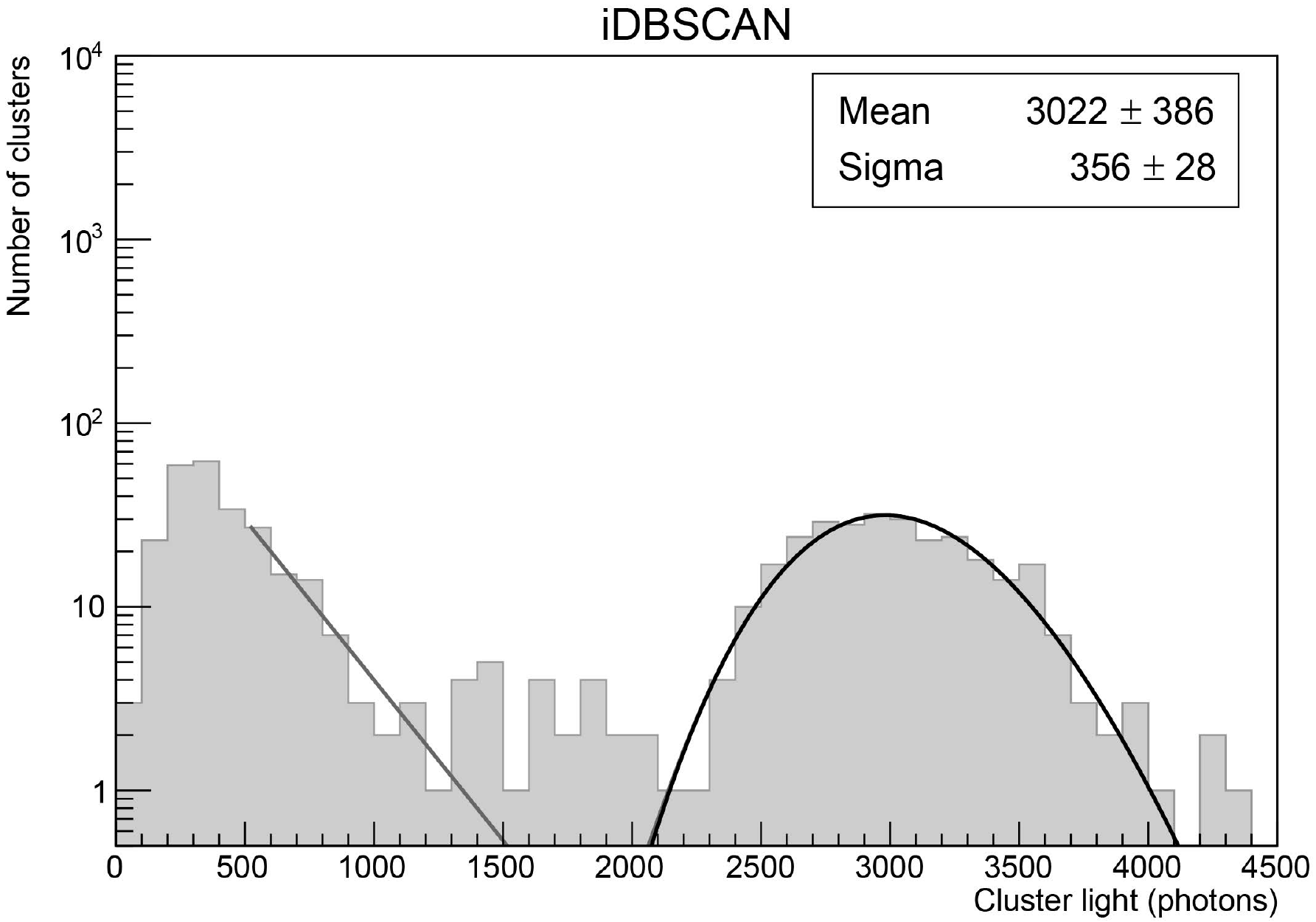}
\caption{Results of the fit applied to the NNC, DBSCAN and iDBSCAN energy distributions for clusters with slimness higher than 0.4.} 
\label{fig_CosFe_slim}
\end{figure}

Figure~\ref{fig_CosFe_slim} shows the fit results when considering only clusters with slimness greater than 0.4.
The estimated energy resolutions are 13.7 $\pm$ 2.4\%, 12.7 $\pm$ 2.3\% and 11.8 $\pm$ 1.7\% for NNC, DBSCAN and iDBSCAN, respectively, with a conversion factor of about 510 ADC units per keV.
Finally, Table~\ref{tab:ResComp} shows the resulting energy resolution for NNC, DBSCAN and iDBSCAN in correspondence of the different thresholds applied to the slimness.
Note that, due to its higher background contamination, the energy resolution obtained with NNC decreases as the slimness threshold value increases, reaching eventually the energy resolution obtained with iDBSCAN.
The energy resolutions obtained with DBSCAN and iDBSCAN are similar and much less dependent on the slimness parameter when compared to NNC, indicating a greater purity in the selection of $^{55}$Fe clusters for these two methods

\begin{table}[ht]
\centering
\caption{Detector resolution comparison between NNC and iDBSCAN as a function of slimness.}
\label{tab:ResComp}
\begin{tabular}{c|cccccc}
\multirow{2}{*}{\begin{tabular}[c]{@{}c@{}}Slimness\\ (width/length)\end{tabular}} & \multicolumn{6}{c}{Resolution (\%)}                                \\
                                                                                   & \multicolumn{2}{c|}{iDBSCAN}   & \multicolumn{2}{c|}{DBSCAN}           & \multicolumn{2}{c}{NNC} \\ \hline \hline
0.0                                                                                & 12.2 & \multicolumn{1}{c|}{$\pm$ 1.8} & 12.6 & \multicolumn{1}{c|}{$\pm$ 2.2} & 18.1    & $\pm$ 4.0    \\
0.2                                                                                & 12.0 & \multicolumn{1}{c|}{$\pm$ 1.7} & 12.6 & \multicolumn{1}{c|}{$\pm$ 2.2} & 17.3    & $\pm$ 3.7    \\
0.4                                                                                & 11.8 & \multicolumn{1}{c|}{$\pm$ 1.8} & 12.7 & \multicolumn{1}{c|}{$\pm$ 2.3} & 13.7    & $\pm$ 2.4    \\
0.6                                                                                & 12.0 & \multicolumn{1}{c|}{$\pm$ 2.0} & 12.9 & \multicolumn{1}{c|}{$\pm$ 2.8} & 11.8    & $\pm$ 1.8    \\
0.8                                                                                & 12.3 & \multicolumn{1}{c|}{$\pm$ 3.8} & 10.4 & \multicolumn{1}{c|}{$\pm$ 3.1} & 11.1    & $\pm$ 2.8   
\end{tabular}
\end{table}


\newpage

\section{Summary}\label{sec:conclusion}

An adapted version of DBSCAN, named intensity-based DBSCAN, has recently been developed and tested on data acquired with a CYGNO TPC prototype. The impact of this algorithm on the detector performance has been studied using 5.9~keV photons from a $^{55}$Fe radioactive source and compared with results obtained with the standard DBSCAN and NNC algorithms.
The iDBSCAN parameters were optimized for the running conditions of LEMOn, which uses a 4M pixels sCMOS camera, and for signals from $^{55}$Fe photons.
The obtained results showed that, with iDBSCAN, the clustering process of the CYGNO's event-reconstruction algorithm can achieve, without any other event-selection routine, a natural radioactivity background rejection in the energy region around 5.9~keV (from 3.0 keV to 8.8 keV) of 0.82$^{+0.04}_{-0.04}$ and a number of electronic-noise clusters per image of $(9 \pm 4)\times 10^{-4}$, occurring predominantly in the region below 1 keV ($\approx$ 500 photons).
Compared to NNC, these results represent an enhancement of 57\% for the former and, for the latter, an improvement by a factor of a few thousand. 
Compared to DBSCAN, iDBSCAN obtained similar performance regarding background rejection in the $^{55}$Fe energy region; however, iDBSCAN has managed to significantly reduce the number of electronic noise clusters also when compared to DBSCAN.
Therefore, despite achieving similar performance in relation to iDBSCAN in the rejection of background radiation, DBSCAN was not as efficient as iDBSCAN in reducing the effects of electronic noise.
Finally, the detector energy resolution using iDBSCAN was measured to be (12.2 $\pm$ 1.8)\% for 5.9 keV electron recoil events. By requiring spots with slimness larger than 0.4, a rate of electronic-noise clusters per image of $(5 \pm 3)\times 10^{-4}$, a natural radioactive background rejection of 0.92$^{+0.03}_{-0.04}$ and an energy resolution of (11.8 $\pm$ 1.7)\% were achieved.

\acknowledgments
This work was supported by the European Research Council (ERC) under the European Union’s Horizon 2020 research and innovation program (grant agreement No 818744) and also by the Coordenação de Aperfeiçoamento de Pessoal de Nível Superior - Brasil (CAPES) - Finance Code 001.

\bibliographystyle{JHEP}
\bibliography{LEMON-20-002}

\end{document}